\documentclass[aps,preprintnumbers,amsmath,amssymb,10pt,floatfix,epsfig,onecolumn,twoside]{revtex4}
\usepackage[pdftex]{graphicx}
\usepackage[usenames, dvipsnames]{color}
\usepackage{amssymb}
\usepackage{bm}
\usepackage{graphicx}
\usepackage{revsymb}
\usepackage{amsmath}
\usepackage{dcolumn}
\usepackage{ulem}

\usepackage{color} 

\newcommand{\mgn}[1]{\textcolor[rgb]{0.5,0.0,0.5}{\bfseries #1}}
\usepackage{xcolor}
\DeclareRobustCommand{\erase}{\bgroup\markoverwith{\textcolor{red}{\rule[.5ex]{2pt}{0.4pt}}}\ULon}

\newcommand{\be}{\begin{eqnarray}}
\newcommand{\ee}{\end{eqnarray}}

\begin{document}
\draft
\title{ Electron-capture rates in the medium-mass nuclei $^{48}$Ti, $^{56}$Ni,
$^{60}$Zn, and $^{64}$Ge within deformed quasiparticle random-phase approximation}

\author{Eunja Ha \footnote{ejaha@hanyang.ac.kr}}
\address{Department of Physics and Research Institute for Natural Science, Hanyang University, Seoul, 04763, Korea}
\author{Myung-Ki Cheoun \footnote{cheoun@ssu.ac.kr}}
\address{Department of Physics and Origin of Matter and Evolution of Galaxies (OMEG) Institute, Soongsil University, Seoul 156-743, Korea}
\author{H. Sagawa \footnote{sagawa@ribf.riken.jp}}
\address{RIKEN, Nishina Center for Accelerator-Based Science, Wako 351-0198, Japan} 
\address{Center for Mathematics and Physics, University of Aizu, Aizu-Wakamatsu, Fukushima 965-8560, Japan}
\address{Institute of Theoretical Physics, Chinese Academy of Sciences, Beijing 100190, China}
\author{Gianluca Col\`o \footnote{colo@mi.infn.it}}
\address{Dipartimento di Fisica, Universit\`a degli Studi, and INFN, Sezione di Milano, via Celoria 16, 20133 Milano, Italy}
\author{Toshitaka~Kajino \footnote{kajino.cosmo.only1@gmail.com}}
\address{School of Physics, Peng Huanwu Collaborative Center for Research and Education, and 
International Research Center for Big-Bang Cosmology and Element Genesis, Beihang 
University, Beijing 100191, China}
\address{Graduate School of Science, The University of Tokyo, Tokyo 113-0033, Japan}
\address{National Astronomical Observatory of Japan, Tokyo 181-8588, Japan}

\begin{abstract}
Electron-capture (EC) rates in medium-mass nuclei are governed by
Gamow--Teller (GT) strength distributions and provide important input for
stellar weak-interaction processes. In this work, we investigate the
deformation dependence of the GT strengths and stellar EC rates in selected
medium-mass nuclei in and near the $pf$ shell, namely $^{48}$Ti, $^{56}$Ni,
$^{60}$Zn, and $^{64}$Ge. The GT$^{(+/-)}$ strength distributions are calculated in the deformed quasiparticle random-phase approximation (DQRPA) on a
single-particle basis obtained by the Skyrme SGII interaction, while the
stellar EC rates are evaluated from the resulting $B(\mathrm{GT}^+)$ strengths
using the standard phase-space formalism. The potential-energy
curves are used to identify shape softness and possible shape coexistence in the nuclei
under consideration. We find that deformation strongly modifies the GT
strength distributions by changing the centroid energies, resonance splitting,
and fragmentation patterns. In particular, a pronounced shape dependence of the GT$^{(+/-)}$ strengths is found for $^{56}$Ni and $^{64}$Ge, whereas $^{60}$Zn is characterized by a favoured
prolate minimum and $^{48}$Ti exhibits a soft near-spherical/prolate
landscape. By contrast, the corresponding EC rates are generally much less
sensitive to deformation than the differential GT response itself, except at
low temperatures and low densities where the low-lying GT$^+$ strength becomes
decisive because of the negative EC $Q$-value in the electron phase space.
Available charge-exchange data for $^{48}$Ti and $^{56}$Ni are used
as benchmarks of the model predictions. The present results provide microscopic
constraints on the role of deformation and shape coexistence in stellar weak
rates for selected medium-mass nuclei, including proton-rich isotopes near the 
$N = Z$ line.
\end{abstract}

\date{\today}

\maketitle

\section{Introduction}

Electron capture (EC) on nuclei is one of the key weak-interaction processes in
stellar environments. In the late stages of massive-star evolution and during
core collapse, EC reduces the electron fraction $Y_e$, weakens the pressure
support from degenerate electrons, and modifies the entropy and composition of
the collapsing core \cite{Langanke2003,Hix2003}. Reliable EC rates of
medium-mass and heavy nuclei are therefore indispensable ingredients in
astrophysical simulations of stellar evolution and supernova dynamics.

EC rates are also relevant on the proton-rich side of the nuclear chart.
Type-I X-ray bursts on accreting neutron stars are powered by the rapid
proton-capture ($rp$) process, which proceeds through sequences of proton
captures and weak decays on nuclei close to the $N=Z$ line
\cite{Wallace1981,Schatz1998,Woosley2004}. Along this path, the reaction flow
can be delayed at waiting-point nuclei, where further proton captures are
hindered by small proton-separation energies and strong photodisintegration.
In particular, $^{64}$Ge, $^{68}$Se, and $^{72}$Kr are well-known waiting-point
nuclei that influence the matter-flow timescale, the burst tail, and the final
nucleosynthesis \cite{Clark2004,Rodriguez2004,Zhou2023}. Under stellar conditions, the effective weak lifetime is not determined by
terrestrial $\beta^+$ decay alone, because continuum EC coming from plasma electrons can contribute
significantly to the total weak-reaction rates and may compete with $\beta^+$ decay at
sufficiently high temperature and density \cite{Sarriguren2009,Sarriguren2011,Chen2024}.

EC is also important in Type Ia supernovae. In particular, in near-Chandrasekhar mass Type Ia supernovae, electron captures on iron-group nuclei increase the
neutronization of the burning matter and strongly affect the final
nucleosynthesis, especially the yields of $^{56}$Ni and neutron-rich iron-group
isotopes \cite{Mori2016,Mori2018}. We also note that Type Ia supernovae are not regarded as a standard site
of the rapid neutron-capture ($r$) process; their nucleosynthesis is dominated instead by thermonuclear
burning to Si-group and Fe-group nuclei, with possible relevance to some
p-nuclide production rather than to the $r$-process nucleosynthesis.

{In Type Ia supernovae, the relevant weak reactions are not the $rp$-process proton
captures but electron captures on the dense thermonuclear ashes,
$A(Z,N)(e^-,\nu_e)B(Z-1,N+1)$. Representative examples are
$p(e^-,\nu_e)n$, $^{56}$Ni$(e^-,\nu_e)^{56}$Co, and
$^{55}$Co$(e^-,\nu_e)^{55}$Fe, with subsequent captures on the daughter nuclei
also possible at sufficiently high density. These reactions lower the electron
fraction $Y_e$ and modify the iron-group composition of the ejecta~\cite{Mori2016,Mori2018}. Their
effect is strongest in high-density near-Chandrasekhar mass explosions and is
therefore useful for distinguishing such models from lower-density burning
scenarios, including white-dwarf mergers.}

A third and more direct motivation comes from the $\nu p$-process in
core-collapse supernovae and hypernovae \cite{Fro2006}. The original motivation for studying
the GT strength functions of $^{56}$Ni, $^{60}$Zn, and $^{64}$Ge is that they
provide key nuclear-structure input for both $(n,p)$ reactions and
charged-current neutrino-induced reactions on these nuclei in the $\nu p$-process. These reactions are
closely related to the long-standing problem of the astrophysical origin of the
relatively abundant p-nuclei $^{92,94}$Mo and $^{96,98}$Ru. But recent studies \cite{Jin2020} 
have suggested that de-excitation of the Hoyle state in $^{12}$C may consume
neutrons, protons, and $\alpha$ particles and thereby drive a strong
$\alpha$-capture flow toward nuclei such as $^{56}$Ni, $^{60}$Zn, and
$^{64}$Ge, potentially suppressing the subsequent $\nu p$-process flow toward
Mo and Ru. In this context, reliable GT distributions and weak-interaction
rates for these nuclei are an essential first step toward constraining the
corresponding $(n,p)$ and neutrino-induced cross sections and, ultimately,
testing whether the hypernova $\nu p$-process can account for the solar-system
abundances of $^{92,94}$Mo and $^{96,98}$Ru \cite{Sasaki2022}.

{In detail, for $^{64}$Ge, the relevant reaction flow depends on the astrophysical
environment. In the $rp$-process, $^{64}$Ge acts as a classical waiting-point
nucleus because its $\beta^+$ half-life is long on the burst timescale, while
the bypass through $^{64}$Ge$(p,\gamma)^{65}$As is hindered by the very small
proton separation energy of $^{65}$As and the strong reverse
$^{65}$As$(\gamma,p)^{64}$Ge channel \cite{Zhou2023}. In the $\nu p$-process, by contrast, the
key escape route is $^{64}$Ge$(n,p)^{64}$Ga, which bypasses the slow $\beta^+$
decay and allows the reaction flow to continue through subsequent proton
captures \cite{Fro2006,Sasaki2022}. Neutron capture, $^{64}$Ge$(n,\gamma)^{65}$Ge, may provide a
secondary path under some conditions, but it is generally less important than
$(n,p)$ for overcoming the $^{64}$Ge bottleneck \cite{Jin2020}. We also note that the $\alpha$-channel is most
important upstream, where a strong $\alpha$-capture flow produces the
$^{56}$Ni--$^{60}$Zn--$^{64}$Ge seeds; if this seed-production flow becomes too
strong, it can deplete the free neutrons, protons, and $\alpha$ particles
required to sustain the subsequent $\nu p$-process.}

The stellar EC rate is governed primarily by the underlying Gamow--Teller
strength in the $\beta^+/{\rm EC}$ direction, $B(\mathrm{GT}^+)$. At low
temperature and density, only a limited phase-space window is available, and
the rate can be controlled by a few low-lying transitions. As the electron
chemical potential increases, higher-lying daughter states become accessible
and the overall location, width, and fragmentation of the GT distribution
become increasingly important \cite{Langanke2003}. Since these features depend
on shell structure, pairing, and deformation, microscopic nuclear-structure
input is essential for reliable EC-rate calculations.

We note that these stellar densities are not fixed by direct laboratory
measurements. Rather, they are inferred from astrophysical models constrained by
observables such as X-ray-burst light curves and by stellar
evolution/core-collapse simulations, while nuclear experiments constrain the
microscopic inputs to the weak rates (masses, $Q$-values, and GT strengths)
rather than the stellar density itself~\cite{Fuller1980,Fuller1982,Oda1994,Langanke2000,Sarriguren2011,Petrovici2015,Woosley2004,Zhou2023,Chen2024}.

Charge-exchange reactions provide valuable benchmarks for GT strengths, but
experimental information remains limited, especially for proton-rich unstable
nuclei \cite{Sasano2011,Fujita2011,Giraud2023}. This makes it necessary to rely on microscopic models capable of
describing deformation and shape competition in a systematic way. In nuclei on
and near the $N=Z$ line, intrinsic deformation, shape softness, and shape
coexistence can strongly modify the low-energy GT response and may therefore
affect the stellar weak-interaction timescale
\cite{Sarriguren2011,Petrovici2015}. Although thermally populated excited
states can also contribute to stellar weak rates, the present work focuses on
the structure dependence of ground-state GT strengths and the corresponding EC
rates.

Here we note that more deliberate calculation of the EC has been performed by a reverse process of the neutrino-induced reaction for  a nucleus, $ A(N,Z)( \nu, e^-) B(N-1,Z+1) $, which included lots of multipole transitions from $0^{\pm}$ to $4^{\pm}$ excitations by the weak interaction through the incident neutrino \cite{Nils2009,Cheoun2010,Fantina2012,Ravic2020}. 
The $0^+\rightarrow1^+$ transition in neutrino-induced reactions is evaluated at finite momentum transfer and therefore includes additional
momentum-dependent contributions of the weak multipole operator. In conventional $\beta$-decay and charge-exchange scattering, the allowed GT transition is defined in the long-wavelength limit ${\bold q} \rightarrow 0$, where the
axial spin operator dominates and the Ikeda sum rule.  In contrast, the $1^+$ strength in the neutrino-induced reaction should not be directly identified with the present conventional GT strength entering the Ikeda sum rule.  

In this work, we study selected medium-mass nuclei in and near the $pf$ shell,
namely $^{48}$Ti, $^{56}$Ni, $^{60}$Zn, and $^{64}$Ge. The nucleus $^{48}$Ti
is included as a benchmark case with available GT$^+$ information, while
$^{56}$Ni, $^{60}$Zn, and $^{64}$Ge probe proton-rich nuclei close to the
$N=Z$ line and exhibit different deformation scenarios. The GT strength
distributions are calculated within a deformed quasiparticle random-phase
approximation (DQRPA), adopting a Skyrme-based quasiparticle basis
obtained with the SGII interaction, and the resulting $B(\mathrm{GT}^+)$ strengths are used
to compute stellar EC rates over representative temperature--density
conditions. In the present work, the stellar EC rates are evaluated from the QRPA GT$^+$ strength
distributions built on the parent ground state. Our main goal is to clarify how deformation and shape coexistence affect the GT spectroscopy and to what extent these structural effects
survive in the integrated EC rates.

The remainder of this paper is organized as follows. Sections~II and III
outline the theoretical framework and the stellar EC-rate formalism.
Section~IV presents the potential-energy curves, GT strength distributions, and
EC rates for the selected nuclei. The main results are summarized and discussed
in the final section.

\section{Formalism}

The present calculations are performed within a DQRPA framework built on a
Skyrme-based quasiparticle basis. The underlying single-particle spectrum and
intrinsic deformation are obtained from an axially deformed Skyrme mean-field
calculation with the SGII interaction in a harmonic-oscillator basis, and
pairing correlations in the quasiparticle vacuum are treated in the Deformed Skyrme-Hartree-Fock-Bogoliubov (DSHFB) framework.
On top of this deformed quasiparticle basis, charge-exchange excitations are
described by the DQRPA using residual particle--hole and particle--particle
interactions derived from a Brueckner $G$-matrix based on the CD-Bonn
nucleon--nucleon potential. The present approach is therefore a hybrid
framework: the static mean field is generated from a Skyrme energy-density
functional, whereas the residual interaction is taken from a realistic
finite-range $NN$ interaction. In the present work, we focus on allowed
Gamow--Teller transitions built on the parent ground state.

To describe GT excitations in axially deformed nuclei, we introduce the usual
DQRPA phonon operator acting on the deformed quasiparticle vacuum. Owing to
axial and reflection symmetries, the projection $K$ of the intrinsic angular
momentum on the symmetry axis and the parity remain good quantum numbers in
the intrinsic frame. The DQRPA excited states are then constructed from
two-quasiparticle proton--neutron configurations with forward and backward
amplitudes determined from the DQRPA equations. In the present implementation,
the static pairing field includes the conventional isovector pairing channel,
whereas explicit isoscalar ($T=0$) pairing correlations are not included.
With these definitions, the phonon creation operator can be written as \cite{Ha2015a}: 
\begin{equation}\label{phonon}
{\cal Q}^{\dagger}_{m,K}  =\sum_{\rho_{\alpha} \alpha \alpha'' \rho_{\beta} \beta \beta''}
[ X^{m}_{( \alpha \alpha'' \beta \beta'')K} A^{\dagger}( \alpha \alpha'' \beta \beta'' K)
- Y^{m}_{( \alpha \alpha'' \beta \beta'')K} {\tilde A}( \alpha \alpha'' \beta \beta'' K)]~,
\end{equation}
{where $\rho_{\alpha (\beta)} (= \pm 1)$ denotes the sign of the total angular momentum projection of the $\alpha$ state for the reflection symmetry.}
Here, we have introduced pair creation and annihilation operators, composed by two quasiparticles and defined as
\begin{equation}
 A^{\dagger}( \alpha \alpha'' \beta \beta'' K)  =
 {[a^{\dagger}_{ \alpha \alpha''} a^{\dagger}_{\beta \beta''}]}^K,
~~~{\tilde A}( \alpha \alpha'' \beta \beta'' K)  =
 {[a_{\beta \beta''} a_{\alpha \alpha''}]}^K,
\end{equation}
where $K$ is the quantum number associated with the projection of the intrinsic angular momentum on the symmetry axis, which is a good quantum number in the axially deformed nuclei.
We note that parity is also treated as a good quantum number in the present approach.  
Here, $\alpha$ indicates a set of quantum numbers to specify the single-particle-state (SPS). Isospin of the real particle is denoted by the Greek letter with prime $(\alpha' , \beta' , \gamma' , \delta')$ (see Eqs. (\ref{eq:mat_A}) and (\ref{eq:mat_B})). Within the quasi-boson approximation for the phonon operator, we obtain the following DQRPA equation for describing the correlated DQRPA ground state:

\begin{eqnarray}\label{qrpaeq}
&&\left(\begin{array}{cccccccc}
           A_{\alpha \beta \gamma \delta (K)}^{1111} & A_{\alpha \beta \gamma \delta (K)}^{1122} &
           A_{\alpha \beta \gamma \delta (K)}^{1112} & A_{\alpha \beta \gamma \delta (K)}^{1121} &
           B_{\alpha \beta \gamma \delta (K)}^{1111} & B_{\alpha \beta \gamma \delta (K)}^{1122} &
           B_{\alpha \beta \gamma \delta (K)}^{1112} & B_{\alpha \beta \gamma \delta (K)}^{1121} \\
           A_{\alpha \beta \gamma \delta (K)}^{2211} & A_{\alpha \beta \gamma \delta (K)}^{2222} &
           A_{\alpha \beta \gamma \delta (K)}^{2212} & A_{\alpha \beta \gamma \delta (K)}^{2221} &
           B_{\alpha \beta \gamma \delta (K)}^{2211} & B_{\alpha \beta \gamma \delta (K)}^{2222} &
           B_{\alpha \beta \gamma \delta (K)}^{2212} & B_{\alpha \beta \gamma \delta (K)}^{2221}\\
           A_{\alpha \beta \gamma \delta (K)}^{1211} & A_{\alpha \beta \gamma \delta (K)}^{1222} &
           A_{\alpha \beta \gamma \delta (K)}^{1212} & A_{\alpha \beta \gamma \delta (K)}^{1221} &
           B_{\alpha \beta \gamma \delta (K)}^{1211} & B_{\alpha \beta \gamma \delta (K)}^{1222} &
           B_{\alpha \beta \gamma \delta (K)}^{1212} & B_{\alpha \beta \gamma \delta (K)}^{1221}\\
           A_{\alpha \beta \gamma \delta (K)}^{2111} & A_{\alpha \beta \gamma \delta (K)}^{2122} &
           A_{\alpha \beta \gamma \delta (K)}^{2112} & A_{\alpha \beta \gamma \delta (K)}^{2121} &
           B_{\alpha \beta \gamma \delta (K)}^{2111} & B_{\alpha \beta \gamma \delta (K)}^{2122} &
           B_{\alpha \beta \gamma \delta (K)}^{2112} & B_{\alpha \beta \gamma \delta (K)}^{2121} \\
             &       &       &      &      &        &           &        \\ \nonumber
           - B_{\alpha \beta \gamma \delta (K)}^{1111} & -B_{\alpha \beta \gamma \delta (K)}^{1122} &
            -B_{\alpha \beta \gamma \delta (K)}^{1112} & -B_{\alpha \beta \gamma \delta (K)}^{1121} &
           - A_{\alpha \beta \gamma \delta (K)}^{1111} & -A_{\alpha \beta \gamma \delta (K)}^{1122} &
           -A_{\alpha \beta \gamma \delta (K)}^{1112}  & -A_{\alpha \beta \gamma \delta (K)}^{1121}\\
           - B_{\alpha \beta \gamma \delta (K)}^{2211} & -B_{\alpha \beta \gamma \delta (K)}^{2222} &
           -B_{\alpha \beta \gamma \delta (K)}^{2212}  & -B_{\alpha \beta \gamma \delta (K)}^{2221} &
           - A_{\alpha \beta \gamma \delta (K)}^{2211} & -A_{\alpha \beta \gamma \delta (K)}^{2222} &
           -A_{\alpha \beta \gamma \delta (K)}^{2212}  & -A_{\alpha \beta \gamma \delta (K)}^{2221}\\
           - B_{\alpha \beta \gamma \delta (K)}^{1211} & -B_{\alpha \beta \gamma \delta (K)}^{1222} &
           -B_{\alpha \beta \gamma \delta (K)}^{1212}  & -B_{\alpha \beta \gamma \delta (K)}^{1221} &
           - A_{\alpha \beta \gamma \delta (K)}^{1211} & -A_{\alpha \beta \gamma \delta (K)}^{1222} &
           -A_{\alpha \beta \gamma \delta (K)}^{1212}  & -A_{\alpha \beta \gamma \delta (K)}^{1221} \\
          - B_{\alpha \beta \gamma \delta (K)}^{2111} & -B_{\alpha \beta \gamma \delta (K)}^{2122} &
           -B_{\alpha \beta \gamma \delta (K)}^{2112}  & -B_{\alpha \beta \gamma \delta (K)}^{2121} &
           - A_{\alpha \beta \gamma \delta (K)}^{2111} & -A_{\alpha \beta \gamma \delta (K)}^{2122} &
           -A_{\alpha \beta \gamma \delta (K)}^{2112}  & -A_{\alpha \beta \gamma \delta (K)}^{2121} \\
           \end{array} \right)\\  && \times
\left( \begin{array}{c}   {\tilde X}_{(\gamma 1 \delta 1)K}^{m}  \\ {\tilde X}_{(\gamma 2 \delta 2)K}^{m} \\
  {\tilde X}_{(\gamma 1 \delta 2)K}^{m} \\  {\tilde X}_{(\gamma 2 \delta 1)K}^{m} \\ \\
     {\tilde Y}_{(\gamma 1 \delta 1)K}^{m} \\ {\tilde Y}_{(\gamma 2 \delta 2)K}^{m} \\
     {\tilde Y}_{(\gamma 1 \delta 2)K}^{m}\\{\tilde Y}_{(\gamma 2 \delta 1)K}^{m} \end{array} \right)
 = \hbar {\Omega}_K^{m}
 \left ( \begin{array}{c} {\tilde X}_{(\alpha 1 \beta 1)K}^{m}  \\{\tilde X}_{(\alpha 2 \beta 2)K}^{m} \\
 {\tilde X}_{(\alpha 1 \beta 2)K}^{m} \\  {\tilde X}_{(\alpha 2 \beta 1)K}^{m}\\ \\
{\tilde Y}_{(\alpha 1 \beta 1)K}^{m} \\ {\tilde Y}_{(\alpha 2 \beta 2)K}^{m} \\
{\tilde Y}_{(\alpha 1 \beta 2)K}^{m} \\ {\tilde Y}_{(\alpha 2 \beta 1)K}^{m} \end{array} \right)  ~,
\end{eqnarray}
where the amplitudes
${\tilde X}^m_{(\alpha \alpha''  \beta \beta'')K }$ and ${\tilde Y}^m_{(\alpha
\alpha''  \beta \beta'')K}$ in Eq. (\ref{qrpaeq}) stand for forward and backward going amplitudes from the state ${ \alpha
\alpha'' }$ to the state  ${\beta  \beta''}$ \cite{Ha2015a} and are related to those in Eq. (\ref{phonon}) by,
$\tilde{X^m}_{(\alpha \alpha'' \beta \beta'')K}=\sqrt2 \sigma_{\alpha \alpha'' \beta \beta''} X^m_{(\alpha \alpha''
 \beta \beta'')K}$
and $\tilde{Y^m}_{(\alpha \alpha'' \beta \beta'')K}=\sqrt2 \sigma_{\alpha \alpha'' \beta \beta''}
Y^m_{(\alpha \alpha'' \beta \beta'')K}$, with a normalization constant $\sigma_{\alpha \alpha'' \beta \beta''}$ = 1 if $\alpha = \beta$ and $\alpha''$ =
$\beta''$, otherwise $\sigma_{\alpha \alpha'' \beta \beta'' }$ = $\sqrt 2$ \cite{Ch93}. 

The $A$ and $B$ matrices in Eq. (\ref{qrpaeq}) are given by
\begin{eqnarray} \label{eq:mat_A}
A_{\alpha \beta \gamma \delta (K)}^{\alpha'' \beta'' \gamma'' \delta''}  = && (E_{\alpha
   \alpha''} + E_{\beta \beta''}) \delta_{\alpha \gamma} \delta_{\alpha'' \gamma''}
   \delta_{\beta \delta} \delta_{\beta'' \delta''}
       - \sigma_{\alpha \alpha'' \beta \beta''}\sigma_{\gamma \gamma'' \delta \delta''}\\ \nonumber
   &\times&
   \sum_{\alpha' \beta' \gamma' \delta'}
   [-g_{pp} (u_{\alpha \alpha''\alpha'} u_{\beta \beta''\beta'} u_{\gamma \gamma''\gamma'} u_{\delta \delta''\delta'}
   +v_{\alpha \alpha''\alpha'} v_{\beta \beta''\beta'} v_{\gamma \gamma''\gamma'} v_{\delta \delta''\delta'} )
    ~V_{\alpha \alpha' \beta \beta',~\gamma \gamma' \delta \delta'}
    \\ \nonumber  &-& g_{ph} (u_{\alpha \alpha''\alpha'} v_{\beta \beta''\beta'}u_{\gamma \gamma''\gamma'}
     v_{\delta \delta''\delta'}
    +v_{\alpha \alpha''\alpha'} u_{\beta \beta''\beta'}v_{\gamma \gamma''\gamma'} u_{\delta \delta''\delta'})
    ~V_{\alpha \alpha' \delta \delta',~\gamma \gamma' \beta \beta'}
     \\ \nonumber  &-& g_{ph} (u_{\alpha \alpha''\alpha'} v_{\beta \beta''\beta'}v_{\gamma \gamma''\gamma'}
     u_{\delta \delta''\delta'}
     +v_{\alpha \alpha''\alpha'} u_{\beta \beta''\beta'}u_{\gamma \gamma''\gamma'} v_{\delta \delta''\delta'})
    ~V_{\alpha \alpha' \gamma \gamma',~\delta \delta' \beta \beta' }],
\end{eqnarray}
\begin{eqnarray} \label{eq:mat_B}
B_{\alpha \beta \gamma \delta (K)}^{\alpha'' \beta'' \gamma'' \delta''}  =
 &-& \sigma_{\alpha \alpha'' \beta \beta''} \sigma_{\gamma \gamma'' \delta \delta''}
  \\ \nonumber &\times&
 \sum_{\alpha' \beta' \gamma' \delta'}
  [g_{pp}
  (u_{\alpha \alpha''\alpha'} u_{\beta \beta''\beta'}v_{\gamma \gamma''\gamma'} v_{\delta \delta''\delta'}
   +v_{\alpha \alpha''\alpha'} v_{{\bar\beta} \beta''\beta'}u_{\gamma \gamma''\gamma'} u_{{\bar\delta} \delta''\delta'} )
   ~ V_{\alpha \alpha' \beta \beta',~\gamma \gamma' \delta \delta'}\\ \nonumber
     &- & g_{ph} (u_{\alpha \alpha''\alpha'} v_{\beta \beta''\beta'}v_{\gamma \gamma''\gamma'}
     u_{\delta \delta''\delta'}
    +v_{\alpha \alpha''\alpha'} u_{\beta \beta''\beta'}u_{\gamma \gamma''\gamma'} v_{\delta \delta''\delta'})
   ~ V_{\alpha \alpha' \delta \delta',~\gamma \gamma' \beta \beta'}
     \\ \nonumber  &- & g_{ph} (u_{\alpha \alpha''\alpha'} v_{\beta \beta''\beta'}u_{\gamma \gamma''\gamma'}
      v_{\delta \delta''\delta'}
     +v_{\alpha \alpha''\alpha'} u_{\beta \beta''\beta'}v_{\gamma \gamma''\gamma'} u_{\delta \delta''\delta'})
   ~ V_{\alpha \alpha' \gamma \gamma',~\delta \delta' \beta \beta'}],
\end{eqnarray}
where the $u_{\alpha \alpha''\alpha'}$ and $v_{\alpha \alpha''\alpha'}$ coefficients are determined from the Hartree Fock Bogoliubov transformation between particles and quasi-particles with isospin $\alpha'$ and $\alpha''$, respectively, in the $\alpha$ state \cite{Ha18-1}. The $g_{\text{pp}}$ and $g_{\text{ph}}$ stand for particle-particle and particle-hole renormalization factors for the residual interactions in Eqs. (\ref{eq:mat_A}) and (\ref{eq:mat_B}).  The two-body interactions $V_{\alpha \beta,~\gamma \delta}$ and $V_{\alpha \delta,~\gamma \beta}$ are particle-particle and particle-hole matrix elements of the residual $N$-$N$ interaction $V$, respectively, which are calculated from the $G$-matrix {as solutions of the Bethe-Goldstone equation from the CD-Bonn potential.}  

{The two-body interactions $V_{\alpha \beta,~\gamma \delta}$ and $V_{\alpha \delta,~\gamma \beta}$ correspond to the $p-p$ and $p-h$ channels, respectively, of the residual $N-N$ interaction in the deformed state. They are 
calculated from the $G$-matrix in the spherical basis as follows
\begin{eqnarray}
V_{\alpha \alpha' \beta \beta' ,~\gamma \gamma' \delta \delta'} 
= - && \sum_{J} \sum_{abcd} F_{\alpha a {\bar\beta} b}^{JK}  F_{\gamma c {\bar \delta} d}^{JK} G( a \alpha' b \beta' c \gamma' d \delta' , J)~, \\ \nonumber
V_{\alpha \alpha' \delta \delta' , \gamma \gamma' \beta \beta'} 
= && \sum_J \sum_{abcd} F_{\alpha a {\delta} d}^{JK}  F_{\gamma
 c {\beta} b}^{JK} G( a \alpha' d \delta' c \gamma' b \beta' , J)~, \\ \nonumber
V_{\alpha \alpha' \gamma \gamma' , \delta \delta' \beta \beta'} 
 = && \sum_J \sum_{abcd} F_{\alpha a {\gamma} c}^{JK}  F_{\beta
 b {\delta} d}^{JK} G( a \alpha' c \gamma' d \delta' b \beta' , J)~.
\end{eqnarray} 
Here $a$ and $\alpha$ indicates spherical and deformed SPS, respectively. {We note that the isospin $\alpha', \beta', \gamma'$ and $\delta'$ in the $G$-matrix is {replaced} by total isospin ($T=0$ or $T=1$) of the two-body interaction in the isospin representation.} We use $F_{\alpha a {\bar \beta} b}^{JK} = B_{a}^{\alpha} B_b^{\beta} C_{j_a \Omega_\alpha j_b \Omega_\beta}$ for $K = \Omega_{\alpha} + \Omega_{\beta}$. The expansion coefficient $B_{\alpha}^a$ is defined as 
\begin{equation}
| \alpha \Omega_{\alpha} > =\sum_{a} B_{a}^{\alpha} |a \Omega_{\alpha} > ~,~  B_{a}^{\alpha} = \sum_{N n_z} C_{l \Lambda { 1 \over 2} \Sigma}^{j \Omega_\alpha} A_{N n_z \Lambda}^{ N_0 l} b_{N n_z \Sigma},  
\end{equation}
with the Clebsch-Gordan coefficient $C_{l \Lambda { 1 \over 2} \Sigma}^{j \Omega_\alpha}$, the spatial overlap integral $A_{N n_z \Lambda}^{ N_0 l}$, and the eigenvalues obtained from the total Hamiltonian in the deformed basis $b_{N n_z \Sigma}$. Detailed formulas regarding the transformation of Eq.~(7) can be found in Ref. \cite{Ha2015a}.}

The GT transition operator ${\hat {\textrm{O}} }_{1\mu}^{\pm}$ is defined by
 \begin{equation} \label{eq:btop}
{\hat {\textrm{O}} }_{1 \mu}^{-}  = \sum_{\alpha \beta }
< \alpha p |\tau^{+} \sigma_K | \beta n >  c_{\alpha p}^{\dagger} {c}_{\beta n},~
{\hat {\textrm{O}}}_{1 \mu}^{+}  =  {( { \hat T}_{1 \mu}^{-} )}^\dagger  = {(-)}^{\mu}
{ \hat {\textrm{O}}}_{1,-\mu}^{-}.
\end{equation}
Detailed calculation for the transformation from the intrinsic frame to the nuclear laboratory system were presented in Ref. \cite{Ha2015a}. 
The ${\hat {\textrm{O}}}^{\pm}$ transition amplitudes from the ground state of an initial (parent) nucleus
to the excited state of a final (daughter) nucleus, {\it i.e.} the one phonon state
$\vert K^+, m\rangle$, are written as
\begin{eqnarray} \label{eq:phonon}
&&< K^+, m | {\hat {\textrm{O}}}_{K }^- | ~QRPA >  \\ \nonumber
&&= \sum_{\alpha \alpha''\rho_{\alpha} \beta \beta''\rho_{\beta}}{\cal N}_{\alpha \alpha''\rho_{\alpha}
 \beta \beta''\rho_{\beta} }
 < \alpha \alpha''p \rho_{\alpha}|  \sigma_K | \beta \beta''n \rho_{\beta}>
 [ u_{\alpha \alpha'' p} v_{\beta \beta'' n} X_{(\alpha \alpha''\beta \beta'')K}^{m} +
v_{\alpha \alpha'' p} u_{\beta \beta'' n} Y_{(\alpha \alpha'' \beta \beta'')K}^{m}], \\ \nonumber
&&< K^+, m | {\hat {\textrm{O}}}_{K }^+ | ~QRPA >  \\ \nonumber
&&= \sum_{\alpha \alpha'' \rho_{\alpha} \beta \beta''\rho_{\beta}}{\cal N}_{\alpha \alpha'' \beta \beta'' }
 < \alpha \alpha''p \rho_{\alpha}|  \sigma_K | \beta \beta''n \rho_{\beta}>
 [ u_{\alpha \alpha'' p} v_{\beta \beta'' n} Y_{(\alpha \alpha'' \beta \beta'')K}^{m} +
v_{\alpha \alpha'' p} u_{\beta \beta'' n} X_{(\alpha \alpha'' \beta \beta'')K}^{m} ]~,
\end{eqnarray}
where $|~QRPA >$ denotes the correlated DQRPA ground state in the intrinsic frame and
the nomalization factor is given as $ {\cal N}_{\alpha \alpha'' \beta
 \beta''} (J) = \sqrt{ 1 - \delta_{\alpha \beta} \delta_{\alpha'' \beta''} (-1)^{J + T} }/
 (1 + \delta_{\alpha \beta} \delta_{\alpha'' \beta''}).$ 

Here, we use the following 
pairing interaction commonly adopted in the SHFB approach, 
\be \label{eq:pairing}
{\cal H}_{pair} ({\bf r}) = {1 \over 2} V_0 [ 1 - V_1 ({\rho / \rho_0  })^{\gamma} ], 
\ee
where the parameters $V_0$, $V_1$ and $\gamma$ are taken from Ref. \cite{Stoitsov}. For the present SGII calculation, the pairing interaction parameters adopted in Eq.~(\ref{eq:pairing}) are listed in Table~I together with the representative deformation parameters selected from the PECs. The pairing energies are obtained with the pairing window associated with an energy cut-off of 60 MeV. In the present implementation, the quasiparticle basis is generated in the DSHFB scheme with the pairing interaction of Eq.~(\ref{eq:pairing}), and we combine a deformed Skyrme mean field with a realistic
Brueckner $G$-matrix residual interaction in the DQRPA framework. In this respect, our approach is not fully self-consistent in the EDF sense.

\begin{table}[t] 
\caption{Pairing-interaction parameters used in the present SGII calculation
together with the representative deformation parameters selected from the DSHFB PECs shown in Fig.~\ref{fig:pec}.}
\begin{ruledtabular}
\begin{tabular}{ccccc}
Nucleus & $V_0$ & $V_1$ & $\gamma$ & adopted $\beta_2$ \\
\hline
$^{48}$Ti & -266.5 & 0.5 & 1.0 & 0.00,\; 0.15 \\
$^{56}$Ni & -266.5 & 0.5 & 1.0 & 0.00,\; 0.36,\; --0.23 \\
$^{60}$Zn & -266.5 & 0.5 & 1.0 & 0.22 \\
$^{64}$Ge & -266.5 & 0.5 & 1.0 & 0.22,\; --0.23 \\
\end{tabular}
\end{ruledtabular}
\end{table}

\section{Stellar electron-capture rate formalism}

We evaluate the stellar electron-capture (EC) rates for the reaction
\begin{equation}
e^- + (Z,A)_i \rightarrow (Z-1,A)_f + \nu_e ,
\end{equation}
where $i$ and $f$ denote the initial and final nuclear states, respectively.
For the temperature--density conditions considered in this work, neutrinos are
assumed to escape freely, and therefore neutrino final-state blocking is neglected.

In a stellar environment, the total EC rate is obtained as a thermal average
over the populated parent states,
\begin{equation}
\lambda = \sum_i P_i \lambda_i ,
\qquad
P_i = \frac{(2J_i+1)\exp(-E_i/kT)}{G} ,
\qquad
G = \sum_i (2J_i+1)\exp(-E_i/kT) ,
\end{equation}
where $J_i$ and $E_i$ are the spin and excitation energy of the parent state,
and $G$ is the nuclear partition function. The rate from a given parent state is
\begin{equation}
\lambda_i = \sum_f \lambda_{if} .
\end{equation}
In the present calculations, however, we restrict ourselves to transitions built
on the parent ground state and therefore use $P_0 = 1$, but $0 \le P_{(i \neq 1)} \le 1$ at finite temperature. 
Within the allowed approximation, the partial EC rate from an initial state $i$
to a final state $f$ can be written in the standard form, 
\begin{equation}
\lambda_{if}
=
\frac{\ln 2}{D} \, B_{if} \, \Phi_{if}(\rho,T) ,
\qquad
D = 6146~\mathrm{s} ,
\end{equation}
where $B_{if}$ is the reduced transition strength and $\Phi_{if}$ is the
phase-space factor. In general,
\begin{equation}
B_{if} = B(F)_{if} + B(\mathrm{GT})_{if} ,
\end{equation}
with
\begin{equation}
B(F)_{if}
=
\frac{1}{2J_i+1}
\left|
\left\langle f \left\| \sum_k \tau_k^\pm \right\| i \right\rangle
\right|^2 ,
B(\mathrm{GT})_{if}
=
\frac{1}{2J_i+1}
\left|
\left\langle f \left\| \sum_k \sigma_k \tau_k^\pm \right\| i \right\rangle
\right|^2 .
\end{equation}
For the nuclei studied here, the EC rates are dominated by allowed
Gamow--Teller transitions, and the numerical calculations are performed from the
microscopic $B(\mathrm{GT}^+)$ strength distributions obtained in the DQRPA. 
Although forbidden transitions are not included in the present calculation,
their contribution is expected to be subleading in the low-density regime where
the EC rates are governed mainly by low-lying GT$^+$ strength. At higher
densities, however, forbidden multipoles may provide non-negligible
quantitative corrections and deserve further investigation. 

In practical calculations of the GT transitions, a commonly adopted
quenching factor is $q \equiv g_A^{\rm eff}/g_A^{\rm free} \approx$ 0.74 in the
$pf$ shell, corresponding to B(GT) $\rightarrow q^2$ B(GT) $\approx 0.55\,$B(GT) \cite{MartinezPinedo1996,Suhonen2017,Caurier2005}. In the present work, the quenching of $g_A$, q = 0.79, from Ref. \cite{Sarri2015}, is employed in all calculations of the $B(GT)$ strengths.

Using the standard dimensionless variables
\begin{equation}
\omega = \frac{E_e}{m_e c^2},
\qquad
p = \sqrt{\omega^2-1},
\qquad
\theta = \frac{kT}{m_e c^2},
\end{equation}
the phase-space factor can be written as
\begin{equation}
\Phi_{if}(\rho,T)
=
\int_{\omega_l}^{\infty}
\omega p \,
(Q_{if}+\omega)^2
F(Z,\omega)
S_e(\omega)
\, d\omega ,
\end{equation}
where $F(Z,\omega)$ is the Fermi function accounting for the Coulomb distortion
of the electron wave function, and $S_e(\omega)$ is the electron occupation
probability. The transition energy entering the neutrino phase space is defined as
\begin{equation}
Q_{if}
=
\frac{Q_{\mathrm{EC}} { -m_e c^2} + E_i - E_f}{m_e c^2} ,
\end{equation}
where $Q_{\mathrm{EC}}$ is the ground-state-to-ground-state EC $Q$ value and
$E_f$ is the excitation energy of the daughter state. The lower integration limit is
\begin{equation}
\omega_l =
\left\{
\begin{array}{ll}
1, & Q_{if} > -1, \\
|Q_{if}|, & Q_{if} < -1 .
\end{array}
\right.
\end{equation}

For the Coulomb correction we use the usual Fermi-function approximation
\begin{equation}
F(Z,\omega)
=
\frac{2\pi\eta}{1-\exp(-2\pi\eta)},
\qquad
\eta = \alpha Z \frac{\omega}{p} ,
\end{equation}
where $\alpha$ is the fine-structure constant. The electron occupation
probability is given by the Fermi--Dirac distribution
\begin{equation}
S_e(\omega)
=
\frac{1}{\exp[(\omega-\mu_e)/\theta] + 1} ,
\end{equation}
where $\mu_e$ is the electron chemical potential in units of $m_e c^2$. The
corresponding positron occupation probability is
\begin{equation}
S_{e^+}(\omega)
=
\frac{1}{\exp[(\omega+\mu_e)/\theta] + 1} .
\end{equation}
The electron chemical potential is determined self-consistently from the stellar
density through
\begin{equation}\label{eq:chemical_potential}
\rho Y_e
=
\frac{1}{\pi^2 N_A}
\left( \frac{m_e c}{\hbar} \right)^3
\int_0^\infty
\left[S_e(\omega)-S_{e^+}(\omega)\right] p^2 dp ,
\end{equation}
where $\rho$ is the baryon density, $Y_e$ is the electron fraction, and $N_A$
is Avogadro's number.

The above expressions provide the basis for the numerical evaluation of the
stellar EC rates once the microscopic $B(\mathrm{GT}^+)$ distribution is known.
In the present work, the rates are calculated from the DQRPA GT strengths on
the parent ground state using experimental $Q_{\mathrm{EC}}$ values and are
presented as functions of temperature and density in Sec.~IV.

\begin{table} 
\caption[bb]{Ground-state-to-ground-state electron-capture $Q_{\mathrm{EC}}$ values
(MeV) used in the phase-space integrals, obtained from experimental nuclear masses \cite{Wang2021}.}
\setlength{\tabcolsep}{2.0 mm}
\begin{tabular}{ccccc}\hline
      Nucleus            & $^{48}$Ti     &  $^{56}$Ni     &  $^{60}$Zn        & $^{64}$Ge       \\ \hline \hline
     $Q_{EC}$[MeV] &  -3.990        &  2.132           &     4.170             & 4.517               \\\hline \hline
 \end{tabular}
\label{tab:12config}
\end{table}

The density variable used in the present EC-rate figures is $\rho Y_e$, which is
the standard stellar weak-rate quantity proportional to the electron number
density, $n_e=N_A\rho Y_e$. In the weak-rate literature, $\rho Y_e$ is therefore
often quoted interchangeably in $g\,\mathrm{cm}^{-3}$ and in
$\mathrm{mol}\,\mathrm{cm}^{-3}$ notation. The range
$\rho Y_e=10^6$--$10^{10}\ \mathrm{mol}\,\mathrm{cm}^{-3}$ adopted here is not
arbitrary. The lower part, $\rho Y_e\sim10^6$--$10^7$, is the most directly
relevant to proton-rich X-ray-burst/$rp$-process environments, for which peak
conditions of $T\sim1$--$3$ GK and $\rho\sim10^6$--$10^7\ \mathrm{g\,cm}^{-3}$
are commonly quoted, and waiting-point weak-rate studies typically consider
$\rho Y_e=10^4$--$10^7\ \mathrm{mol}\,\mathrm{cm}^{-3}$. The higher densities,
$10^8$--$10^{10}$, are included to display the systematic density dependence of
the rates and to facilitate comparison with the classical FFN/Oda/Langanke
stellar weak-rate grids and with presupernova/core-collapse studies, where
$\rho Y_e\sim10^7$--$10^{10}\ \mathrm{mol}\,\mathrm{cm}^{-3}$ is standard~\cite{Fuller1980,Fuller1982,Oda1994,Langanke2000}.
Thus, for the present proton-rich nuclei, the panels at
$\rho Y_e=10^6$ and $10^7\ \mathrm{mol}\,\mathrm{cm}^{-3}$ are the most directly
relevant to $rp$-process applications, whereas the
$\rho Y_e=10^8$--$10^{10}\ \mathrm{mol}\,\mathrm{cm}^{-3}$ panels should be
viewed as an extension to the broader stellar weak-rate regime.

\section{Results and discussion}
\begin{figure} 
\includegraphics[width=0.45\linewidth]{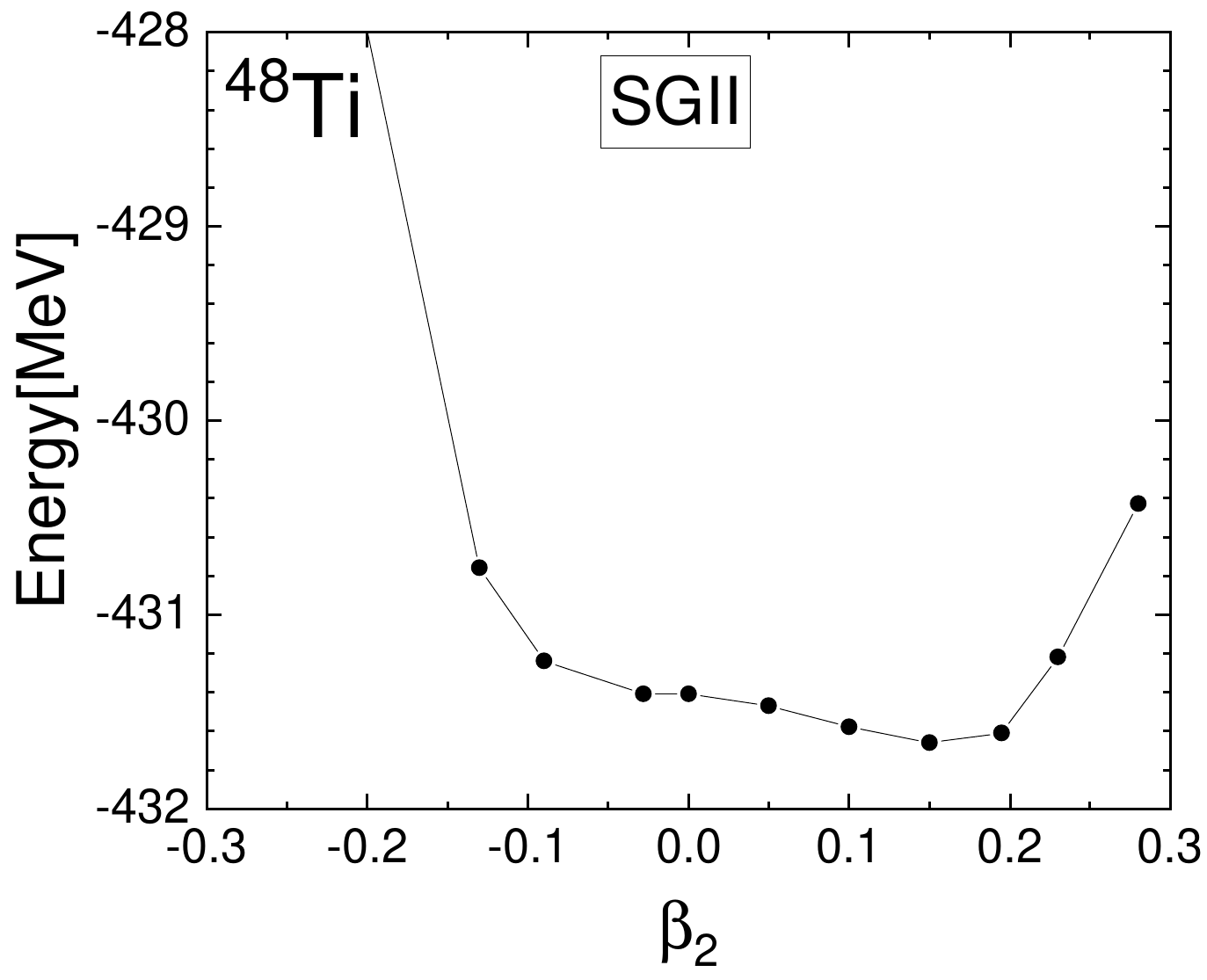}
\includegraphics[width=0.45\linewidth]{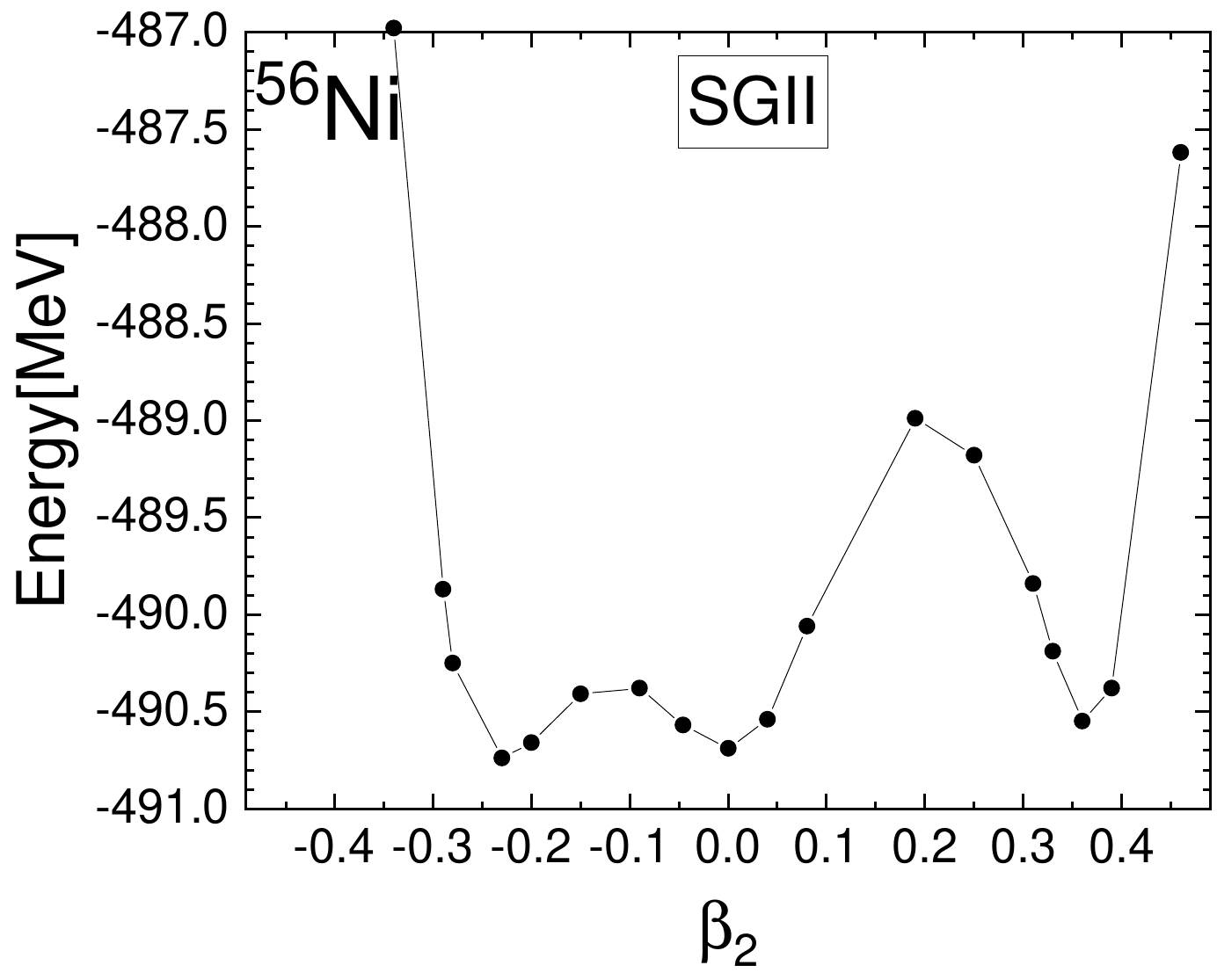}
\includegraphics[width=0.45\linewidth]{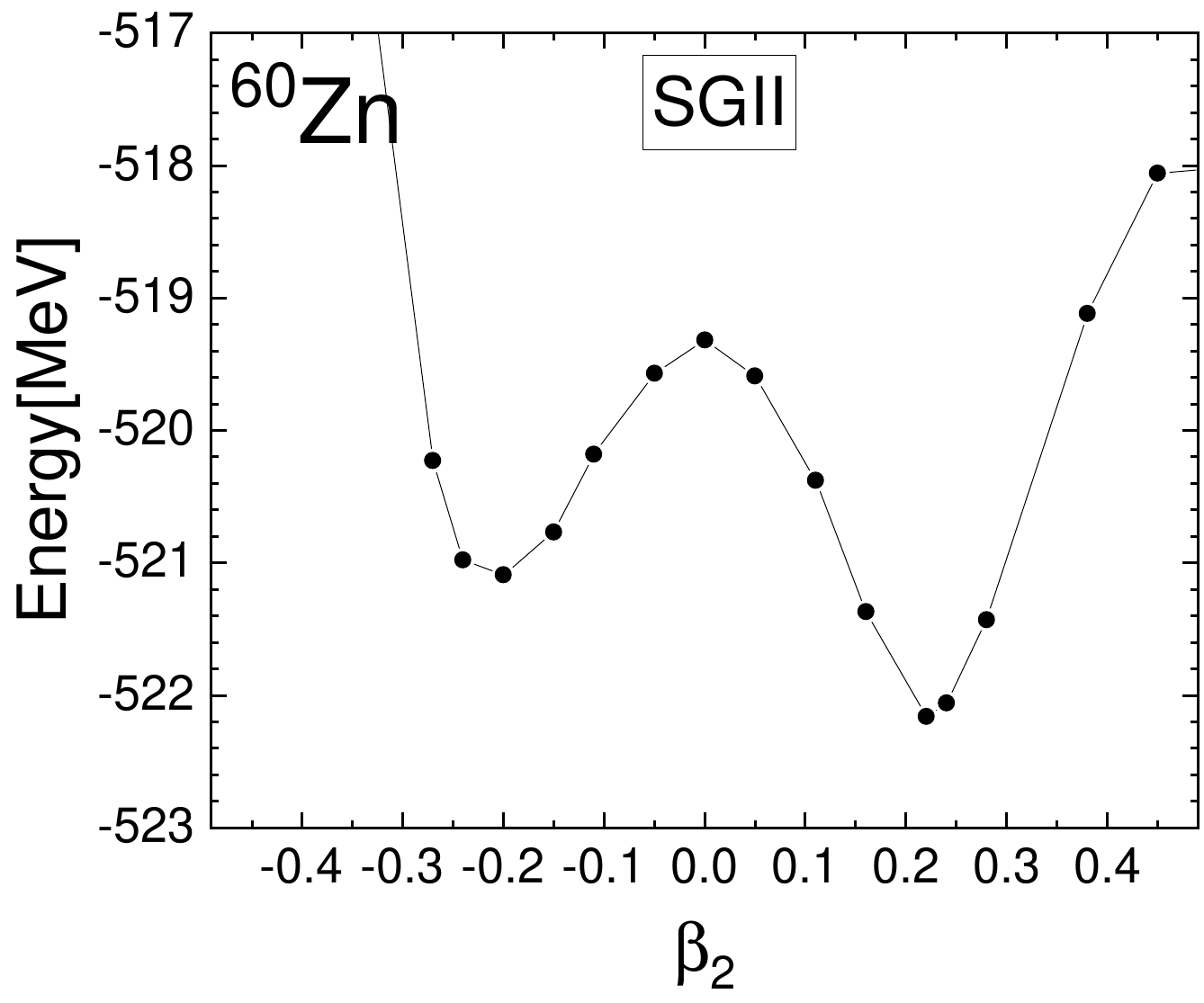}
\includegraphics[width=0.45\linewidth]{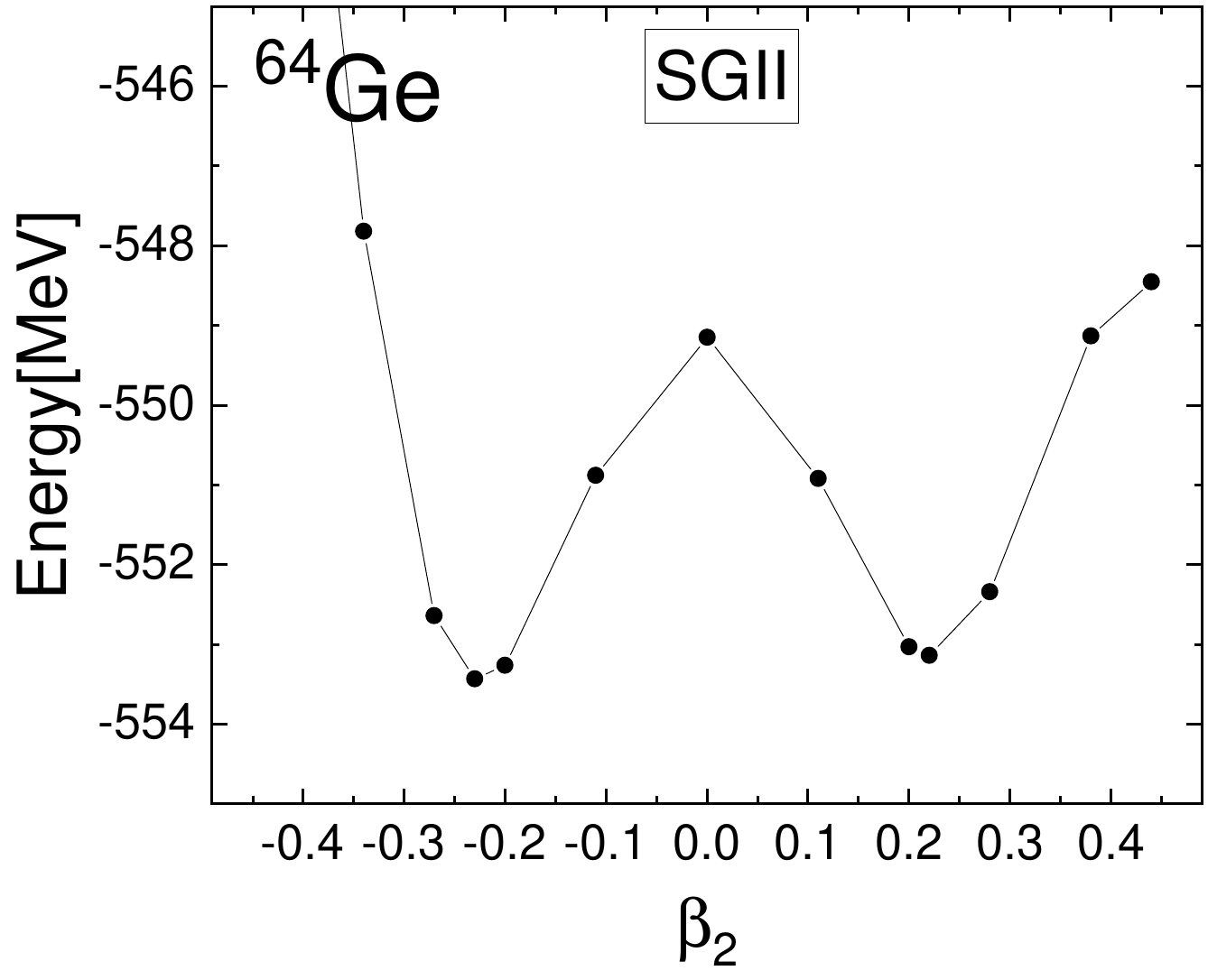}
\caption{(Color online) {Potential energy curves (PECs) of $^{48}$Ti, $^{56}$Ni, $^{60}$Zn, and $^{64}$Ge obtained with the SGII
interaction as a function of the quadrupole deformation parameter $\beta_2$. Negative (positive) $\beta_2$ values correspond to oblate (prolate) shapes. The presence of multiple local minima indicates competing intrinsic configurations and suggests 
softness or shape coexistence.}}
\label{fig:pec}
\end{figure}

In this section, we present the SGII results for the potential-energy curves
(PECs), Gamow--Teller strength distributions, and stellar electron-capture
rates of $^{48}$Ti, $^{56}$Ni, $^{60}$Zn, and $^{64}$Ge. The representative
intrinsic shapes adopted in the GT and EC calculations are selected from the
PECs shown in Fig.~\ref{fig:pec}, and the corresponding deformation parameters
are summarized in Table~I. For each nucleus, deformation values corresponding to local minima within an energy difference of 0.3 MeV are selected. The ground-state-to-ground-state
$Q_{\mathrm{EC}}$ values used in the phase-space integrals are listed in
Table~II. We first discuss the deformation landscape of each nucleus and then
examine how it is reflected in the GT response and in the resulting
electron-capture rates under stellar temperature--density conditions.

The SGII PECs in Fig.~\ref{fig:pec} display distinct deformation patterns in
the nuclei considered here. The nucleus $^{48}$Ti exhibits a relatively soft
prolate minimum around $\beta_2\approx0.15$, while the spherical
configuration lies about 0.25 MeV higher in energy, indicating shape softness rather than a
rigidly deformed structure. In $^{56}$Ni, the PEC shows competing spherical,
prolate, and oblate configurations within the present axial mean-field
framework. For $^{60}$Zn, the lowest minimum is clearly prolate, whereas the
oblate minimum remains higher in energy by about 1.07 MeV. 
Finally, $^{64}$Ge exhibits nearly degenerate prolate and oblate minima, with the prolate one lying about 0.29 MeV higher in energy,
suggesting pronounced oblate--prolate shape coexistence or strong shape softness. These different deformation
landscapes provide the structural basis for the GT and EC systematics discussed below.

%
\begin{figure} 
\includegraphics[width=0.45\linewidth]{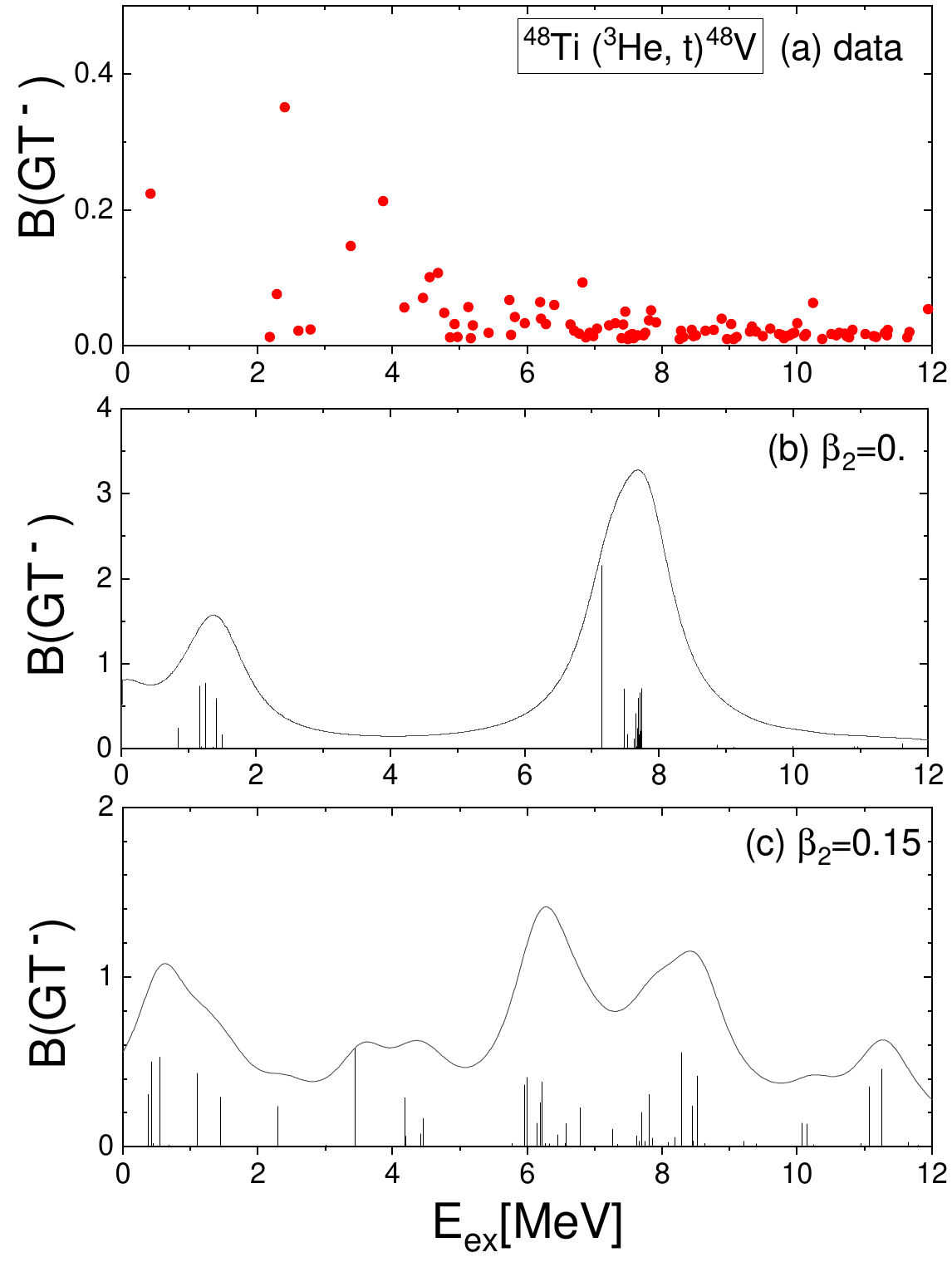}
\includegraphics[width=0.45\linewidth]{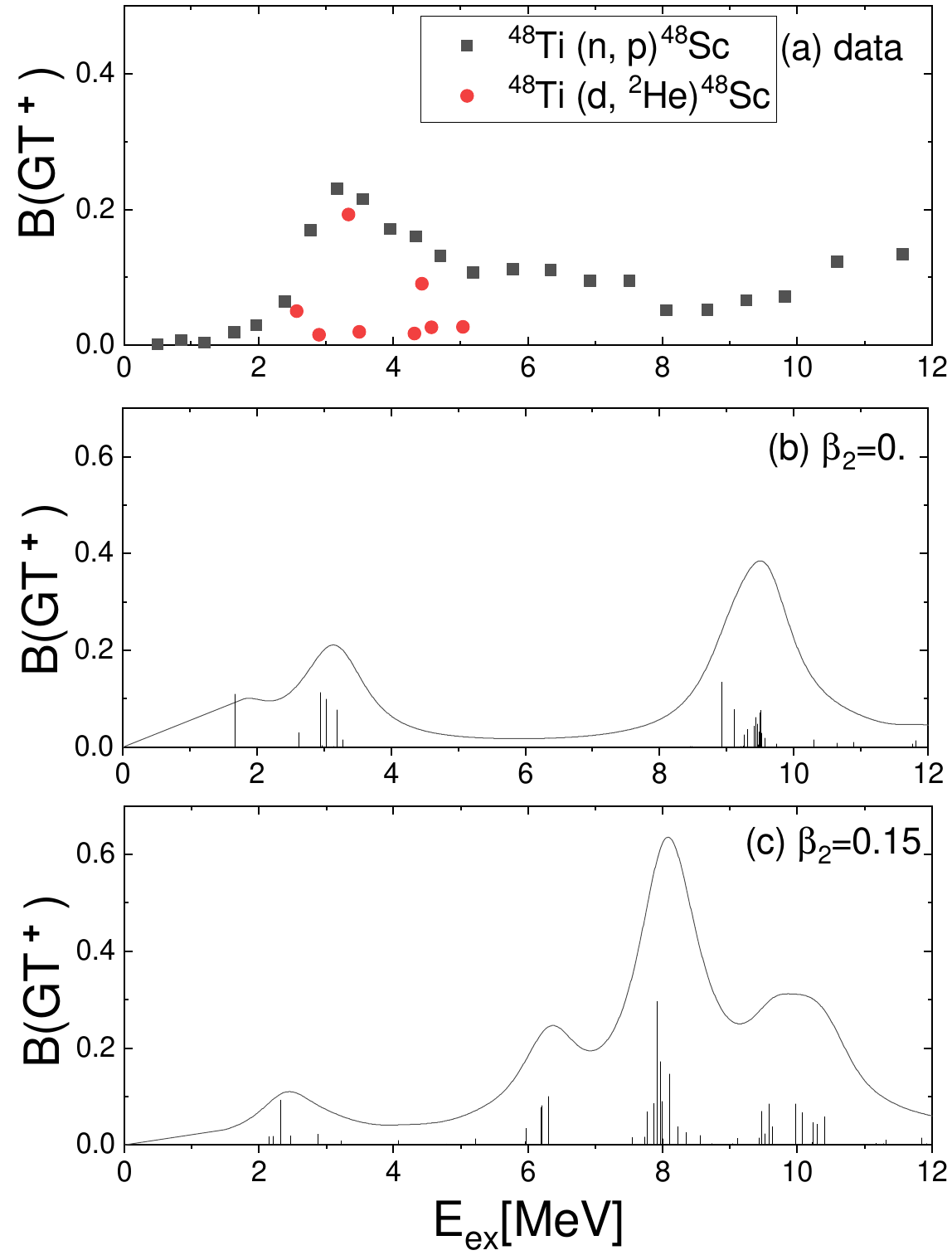}
\caption{(Color online)
 Gamow--Teller $B(\mathrm{GT}^-)$ and $B(\mathrm{GT}^+)$ strength
distributions for $^{48}$Ti calculated with the SGII interaction. The left and
right columns correspond to the $^{48}\mathrm{Ti}\rightarrow{}^{48}\mathrm{V}$
GT$^-$ and $^{48}\mathrm{Ti}\rightarrow{}^{48}\mathrm{Sc}$ GT$^+$ channels,
respectively. In panel (a) of the left column, the experimental GT$^-$ data are
taken from the $(^3\mathrm{He},t)$ reaction ~\cite{Ganio2016}. In panel (a) of the right
column, the experimental GT$^+$ data are taken from the $(n,p)$ and
$(d,{}^{2}\mathrm{He})$ reactions ~\cite{Yako09,Rakers04}. Panels (b) and (c) display the
theoretical results for the constrained shapes $\beta_2=0$ and $\beta_2=0.15$,
respectively. A quenching factor $q=0.79$ is introduced in the B(GT) calculations. For visual comparison with the experimental distributions, in the whole GT results, the discrete
GT strengths were folded by the Lorentzian function with a full width at half maximum (FWHM) of 1.0 MeV. 
}
\label{fig:48Ti_gtmp}
\end{figure}

\begin{figure} 
\includegraphics[width=0.45\linewidth]{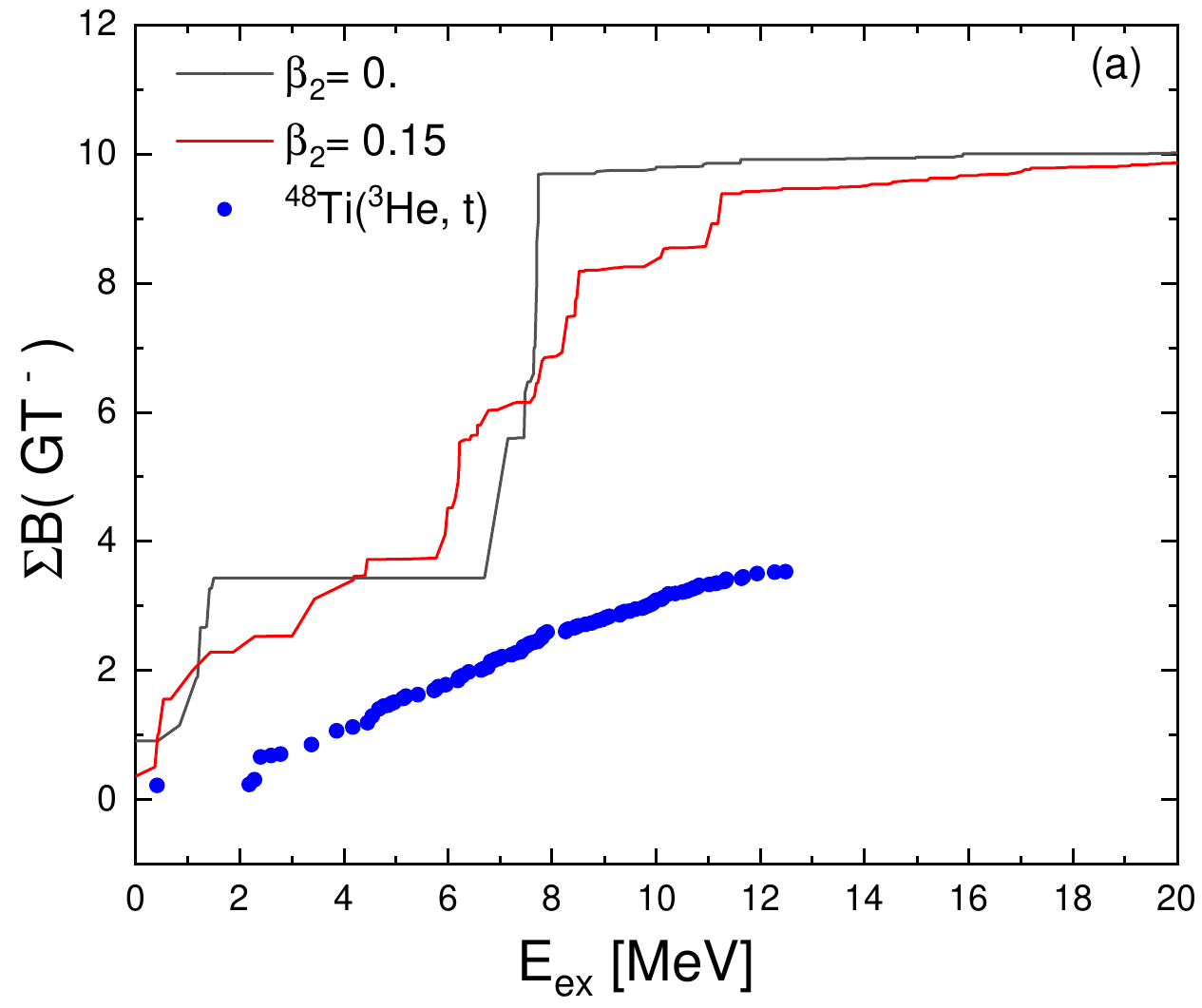}
\includegraphics[width=0.45\linewidth]{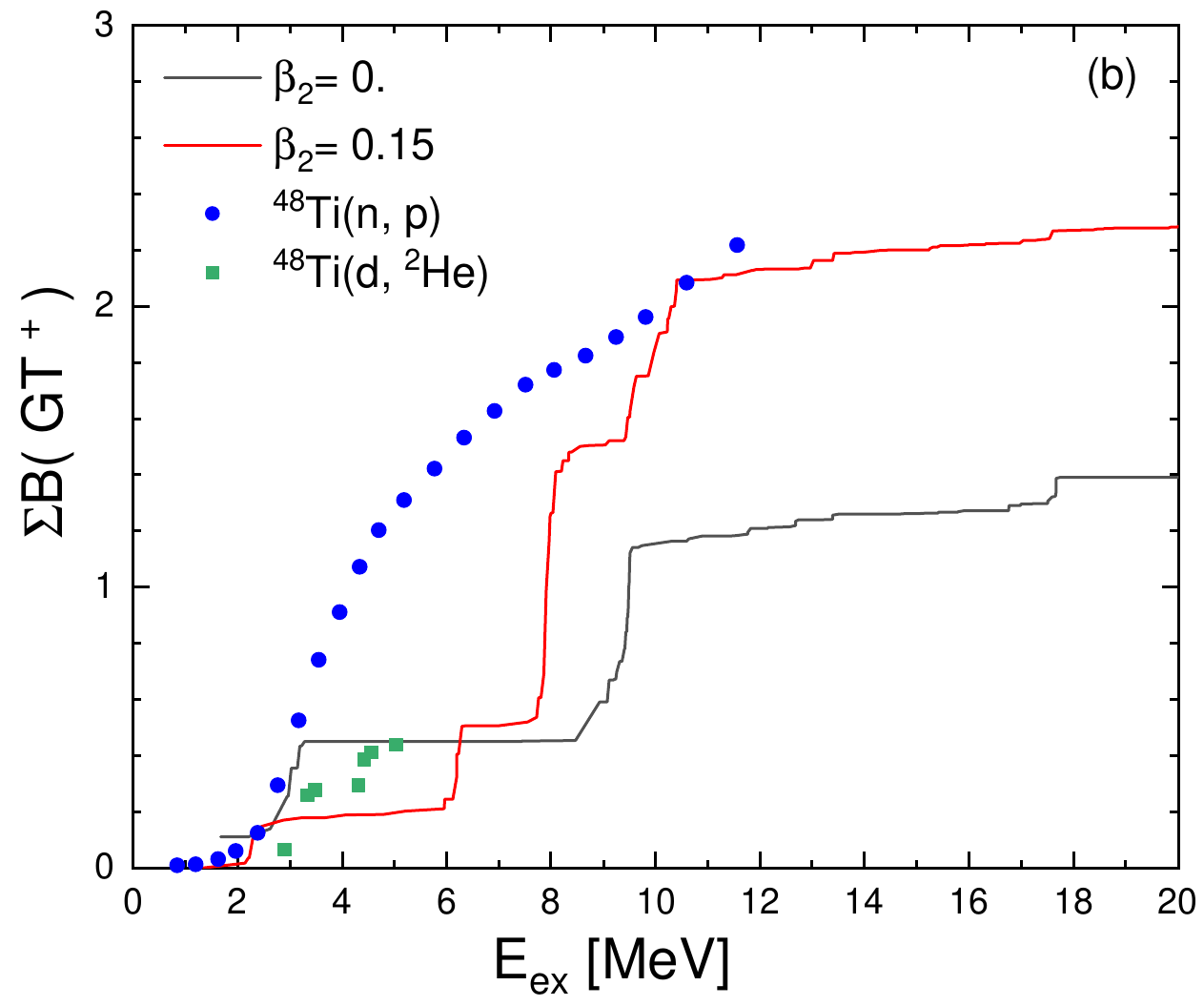}
\caption{(Color online) The \mgn{cumulative} sum of \mgn{the} $B(\mathrm{GT}^-)$ and $B(\mathrm{GT}^+)$ strength
distributions for $^{48}$Ti calculated with the SGII interaction. The left and 
right columns correspond to the $^{48}\mathrm{Ti}\rightarrow{}^{48}\mathrm{V}$
GT$^-$ and $^{48}\mathrm{Ti}\rightarrow{}^{48}\mathrm{Sc}$ GT$^+$ channels,
respectively. In panel (a), the experimental GT$^-$ data are
taken from the $(^3\mathrm{He},t)$ reaction ~\cite{Ganio2016}. In panel (b), the experimental GT$^+$ data are taken from the $(n,p)$ and
$(d,{}^{2}\mathrm{He})$ reactions ~\cite{Yako09,Rakers04}. In each panel, theoretical results for the constrained shapes $\beta_2=0$ and $\beta_2=0.15$ are displayed.}
\label{fig:48Ti_gtmp}
\end{figure}

%
\begin{figure} 
\includegraphics[width=0.47\linewidth]{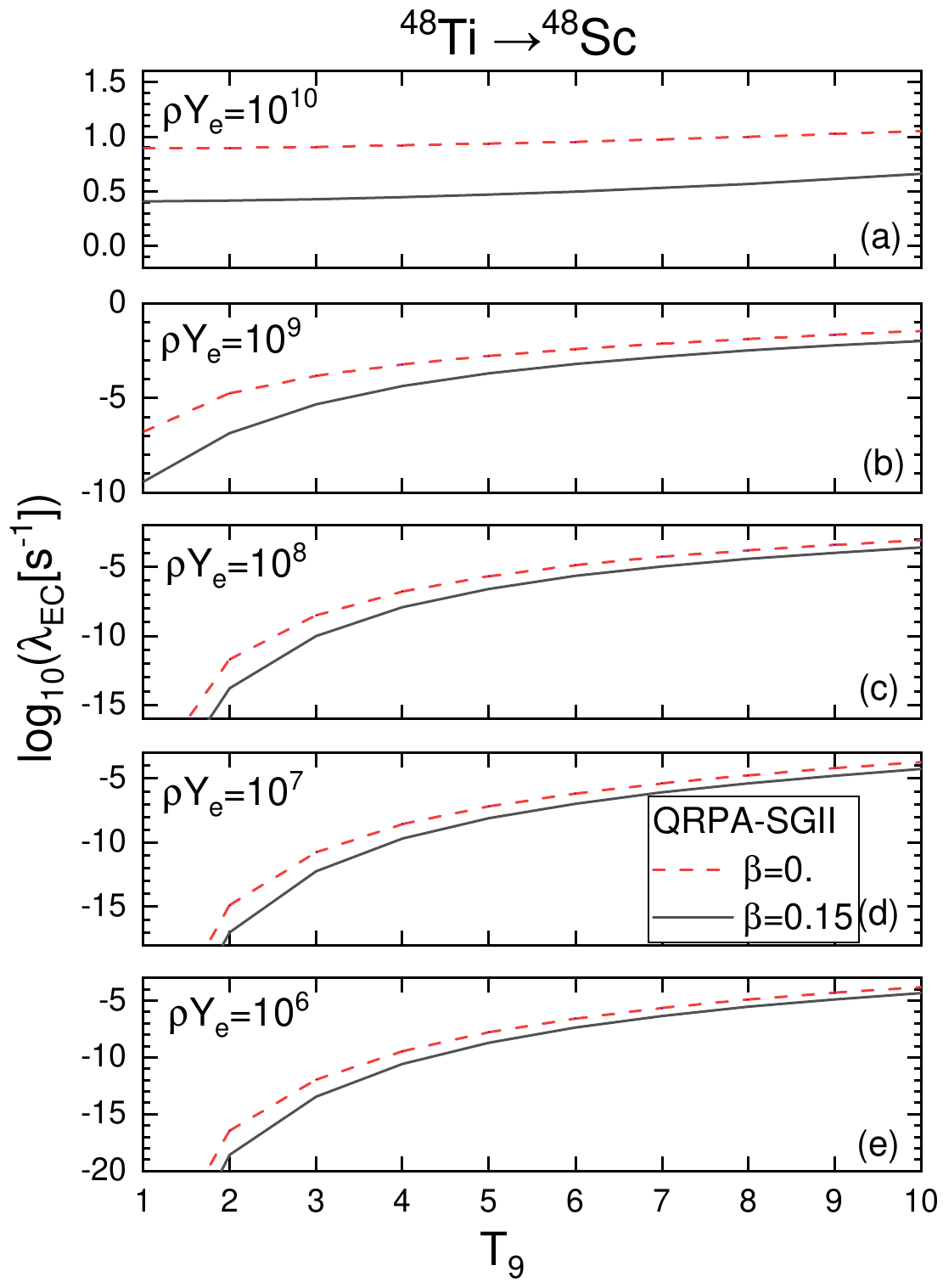}
\includegraphics[width=0.47\linewidth]{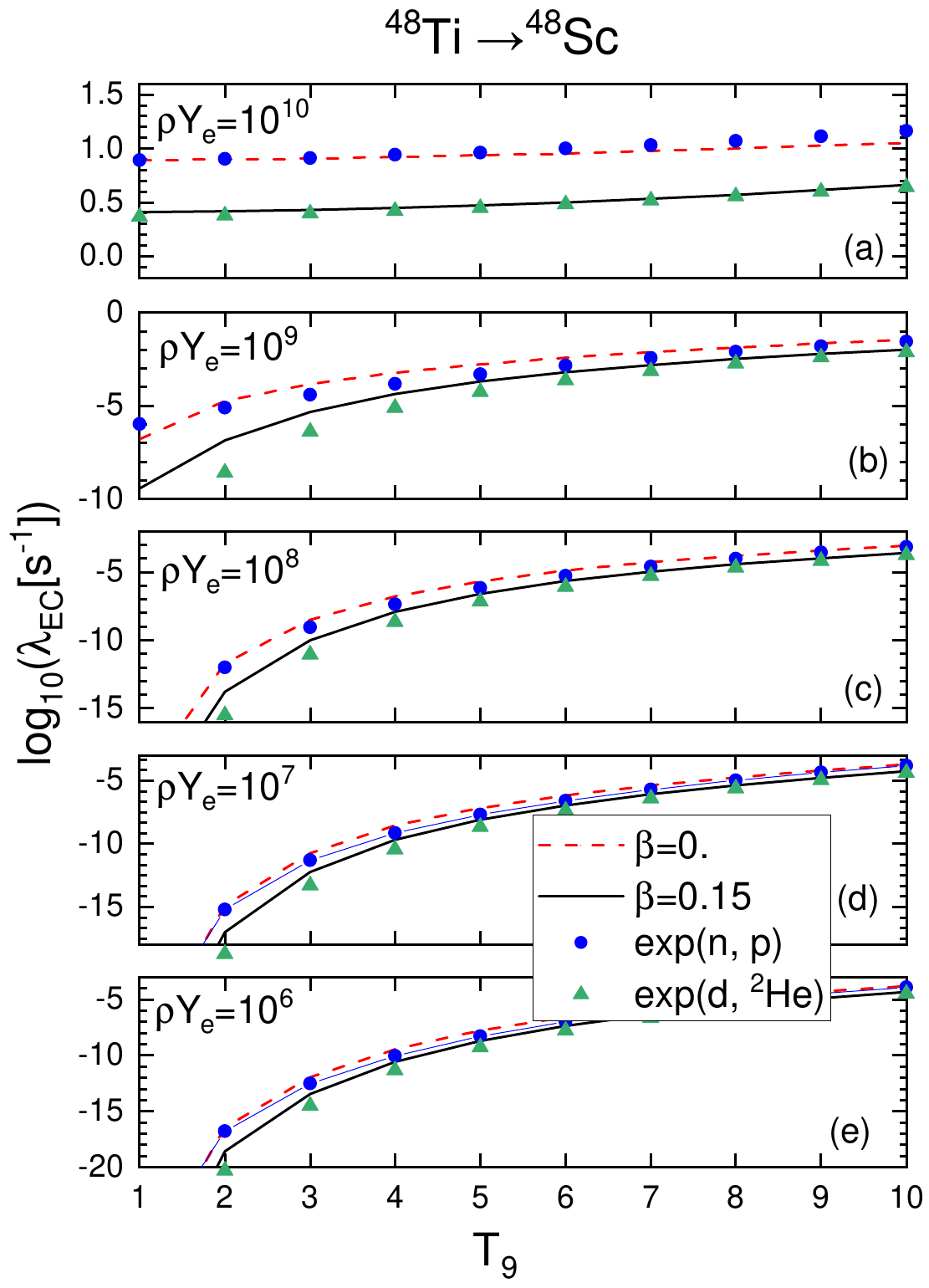}
\caption{({Color online) Electron-capture rates for $^{48}\mathrm{Ti}$ as a function of
the temperature $T_9 \equiv 10^9 \times T$ (Kelvin) calculated from the GT$^+$ strength distributions shown in
Fig.~\ref{fig:48Ti_gtmp}. The left column corresponds to the results
obtained for the constrained shapes $\beta_2=0$ and $\beta_2=0.15$,
respectively. In the right panel, the theoretical rates are compared with the rates
derived from the experimental GT$^+$ strengths extracted from the $(n,p)$ and
$(d,{}^{2}\mathrm{He})$ data. From top to bottom, the densities are
$\rho Y_e = 10^{10}, 10^9, 10^8, 10^7,$ and $10^6$ mol/cm$^3$.}}
\label{fig:48Ti_rate}
\end{figure}
\subsection{GT strength and electron-capture rates in $^{48}$Ti}
\label{subsec:gt_ec_48ti}
The  energy surface of $^{48}$Ti shown in Fig.~\ref{fig:pec} suggests 
softness, or even a possible shape
coexistence between near-spherical and weakly prolate
configurations, rather than two well-separated strongly deformed minima.
Accordingly, the comparison between the constrained solutions at $\beta_2=0$
and $\beta_2=0.15$ provides a physically meaningful test of the sensitivity of
spin-isospin properties to the intrinsic nuclear shape.

The B(GT$^-$) data for $^{48}$Ti in the left panels of Fig.~\ref{fig:48Ti_gtmp} indicate a fragmented low-energy strength distribution rather
than a single concentrated resonance. The spherical calculation overestimates the collectivity
around $E_{\mathrm{ex}}\approx 7$ MeV, whereas the deformed solution improves the fragmentation
but shifts too much strength to higher excitation energies. Therefore, the GT$^-$ response of
$^{48}$Ti supports the interpretation of this nucleus as a shape-soft system rather than a rigid
spherical or rigidly deformed one.

In Fig.~3, the cumulative sum of GT$^-$ and GT$^+$ strengths are displayed by taking into account the quenching factor, $q$ = 0.79. 
In the left panel, the calculated results for the  cumulative sum of the $B(GT^-)$ strength are approximately a factor 2 larger than the experimental values, 
while the GT energy positions in Fig.~2 are reasonably explained.  
The results of the shell model calculation show similar overestimation of about 1.5 times, which was mentioned in connection with the experimental data in Ref. \cite{Ganio2016}.  The shell model results were obtained by using the GXPF1J interaction \cite{Honma2004}, $q$ = 0.74, and the restricted particle model space up to the $pf$ shell with an inert core $^{40}$Ca. 
We notice also that  the experimental $B(GT^{-})-B(GT^{+})$ value up to $E_{ex} =$ 12 MeV is less than 1, which  departs largely from  the Ikeda sum rule, $3(N-Z) = 12$.  
This fact may hint to a large missing strength in the experimental observations.

The right panels in Fig.~\ref{fig:48Ti_gtmp} displays the calculated $B(\mathrm{GT}^+)$ strength
distributions for $^{48}$Ti obtained with the SGII interaction, together with
the available experimental data from the $(n,p)$ and $(d,{}^{2}\mathrm{He})$
reactions. In the spherical solution, part of the GT$^+$ strength remains in
the low-lying region below $E_{\mathrm{ex}}\approx 5$ MeV, while another
fragmented component appears around $E_{\mathrm{ex}}\approx9$--$10$ MeV. In
contrast, for the deformed solution at $\beta_2=0.15$, a substantial fraction
of the strength is shifted to higher excitation energies, with a more
pronounced concentration in the $E_{\mathrm{ex}}\approx7$--$10$ MeV region.
Therefore, the main effect of deformation is not simply to increase or reduce
the total GT$^+$ strength, but rather to redistribute it and move its centroid
to higher excitation energies. The comparison with experiment also shows that
the empirical GT$^+$ strength itself is spread over a broad energy range and
that noticeable differences exist between the two experimental datasets,
especially in the low-energy region most relevant for stellar weak rates.
These differences likely reflect, at least in part, the different extraction
procedures and experimental sensitivities of the two charge-exchange probes. 

The corresponding electron-capture rates are shown in
Fig.~\ref{fig:48Ti_rate}. At high densities, where the electron chemical
potential is sufficiently large to access a wide range of final states, the
calculated rates for $\beta_2=0$ and $\beta_2=0.15$ differ only slightly as shown in the left panels. In
this regime, the details of the GT$^+$ fragmentation are largely averaged out
by the available phase space. On the other hand, at low densities and low
temperatures, the capture rates become strongly dependent on the low-lying
GT$^+$ strength and therefore show enhanced sensitivity to both the adopted
experimental input and the intrinsic shape. Even in this regime, however, the
difference between the rates derived from the two experimental GT datasets is
comparable to, or larger than, the change induced by the moderate static
deformation considered here. These results indicate that, for $^{48}$Ti, the
electron-capture rates are relatively robust against moderate deformation under
medium- and high-density stellar conditions, whereas the low-density regime is
controlled primarily by the detailed distribution of the low-energy GT$^+$
strength.

%

\subsection{GT strengths and electron-capture rates in $^{56}$Ni}
\label{subsec:gt_ec_56ni}

\begin{figure} 
\includegraphics[width=0.45\linewidth]{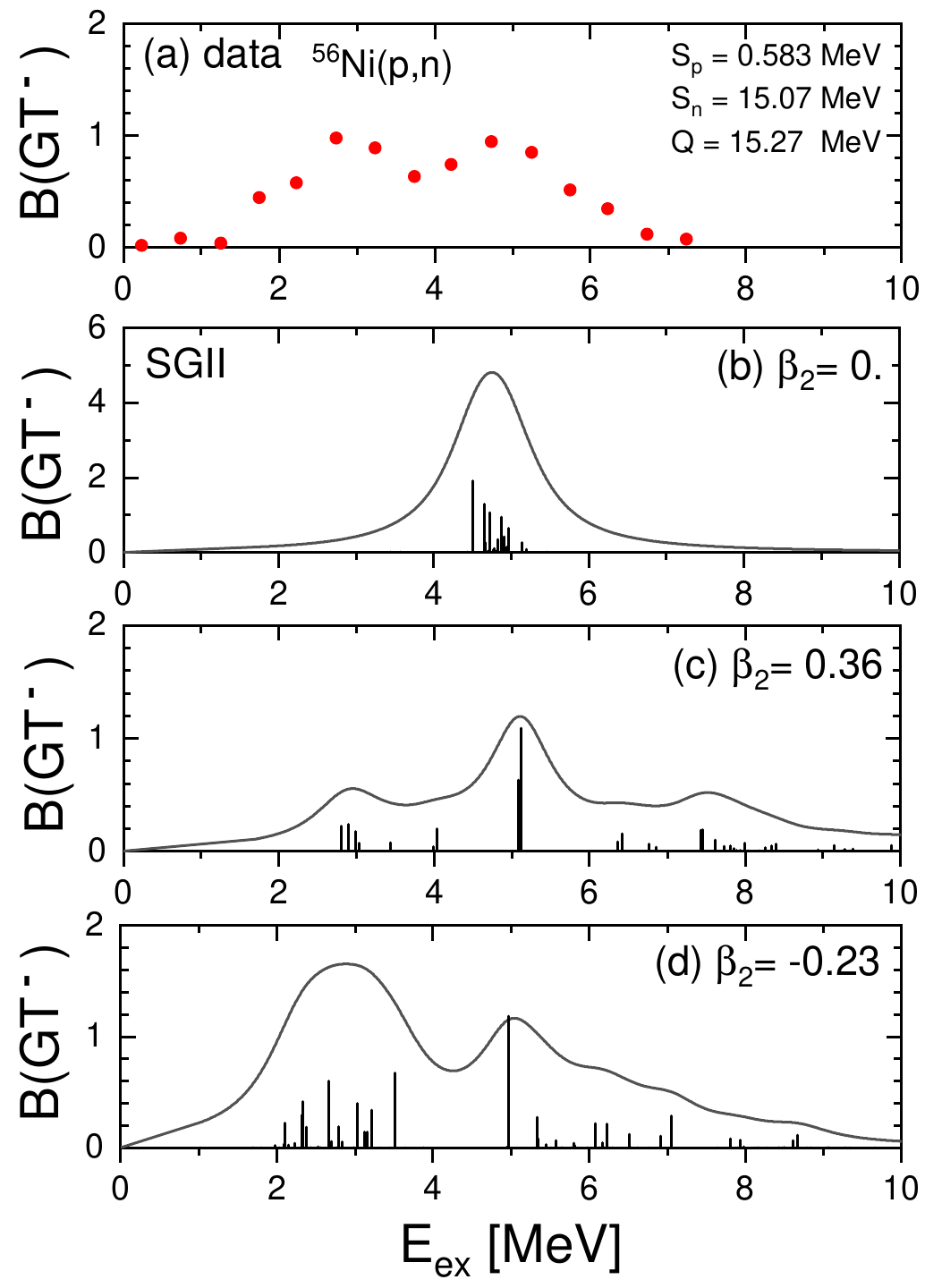}
\includegraphics[width=0.45\linewidth]{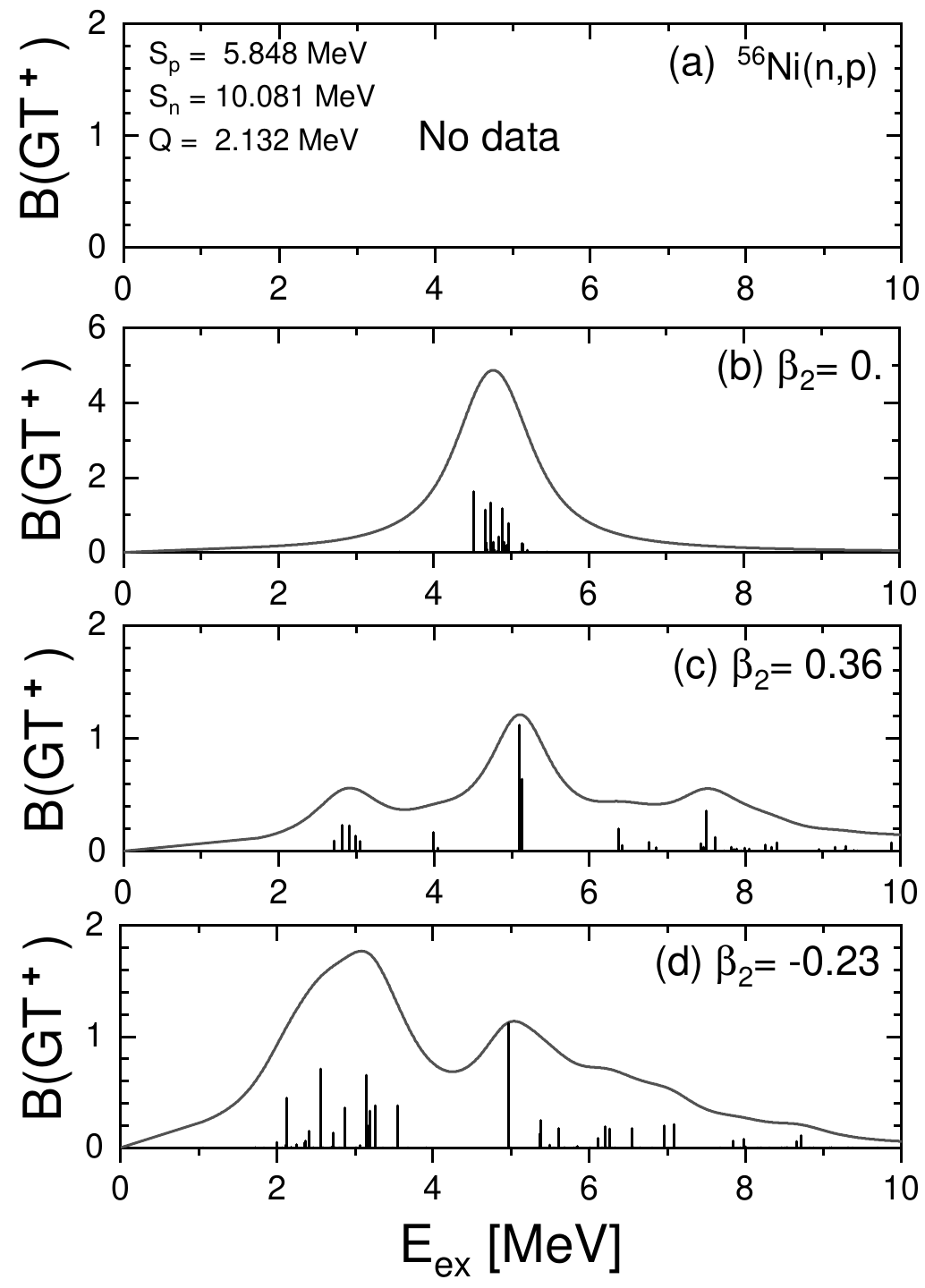}
\caption{(Color online) Gamow--Teller $B(\mathrm{GT}^-)$ and $B(\mathrm{GT}^+)$ strength
distributions in $^{56}\mathrm{Ni}$ calculated with the SGII interaction. The
left and right columns correspond to the $\mathrm{GT}^-$ and $\mathrm{GT}^+$
channels, respectively. Panel (a) shows the available experimental
$^{56}\mathrm{Ni}(p,n)$ data for the $\mathrm{GT}^-$ channel, while no direct
experimental data are available for the $\mathrm{GT}^+$ channel. Panels (b),
(c), and (d) display the theoretical results obtained for the constrained
shapes $\beta_2=0$, $\beta_2=0.36$, and $\beta_2=-0.23$, respectively.
Experimental data are taken from Ref.~\cite{Sasano2011}.
}
\label{fig:56Ni_gtmp}
\end{figure}
%

\begin{figure} 
\includegraphics[width=0.75\linewidth]{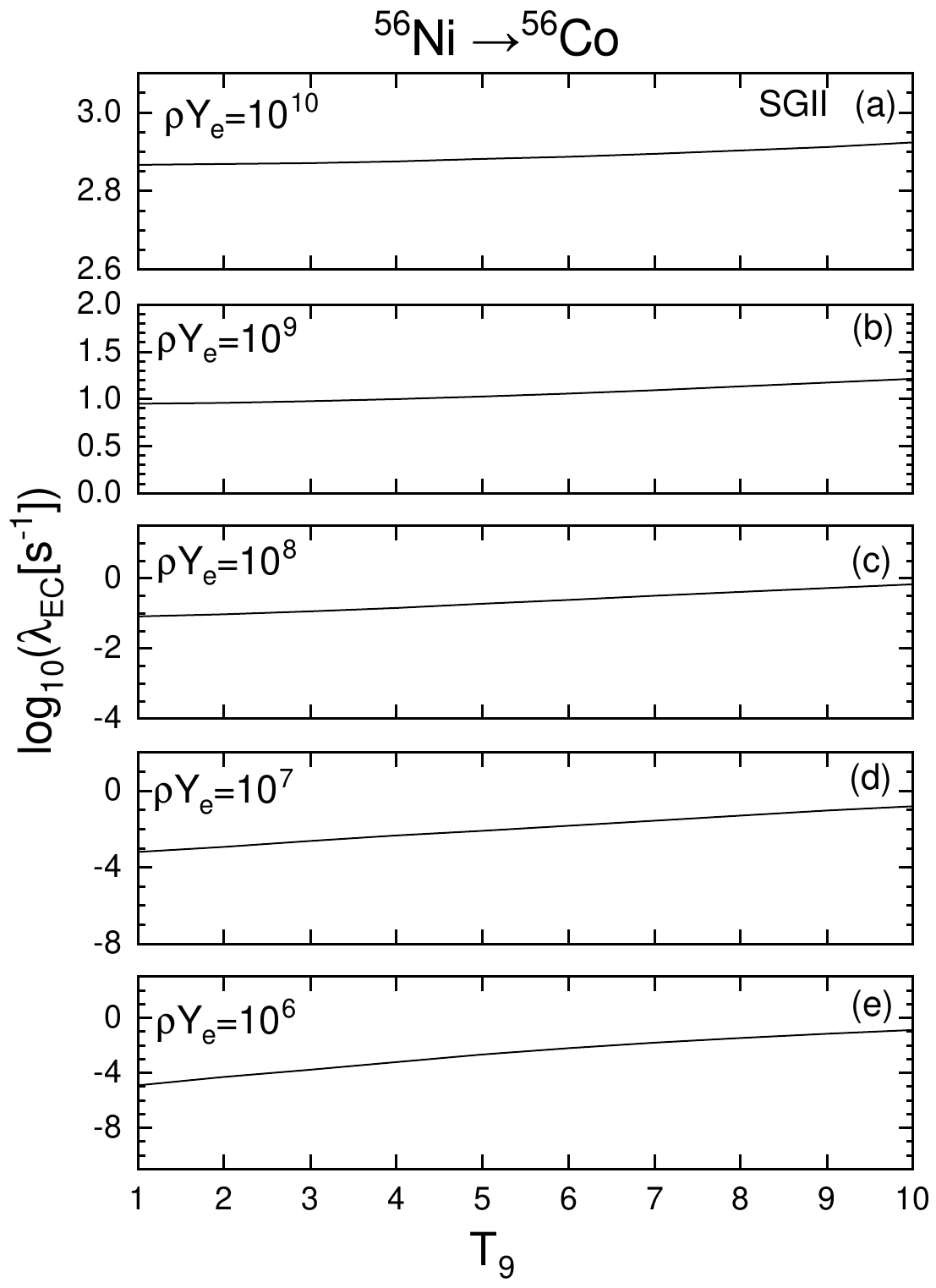}
\caption{(Color online) Same as Fig.~\ref{fig:48Ti_rate}, but for $^{56}\mathrm{Ni}$.
The results for the competing shapes are nearly indistinguishable on the scale
of the figure, indicating a weak shape dependence of the electron-capture
rates.}
\label{fig:56Ni_rate}
\end{figure}
Since the potential-energy curve
of $^{56}$Ni indicates pronounced shape coexistence among the spherical and oblate configurations, the constrained solutions at
$\beta_2=0$, $\beta_2=0.36$, and $\beta_2=-0.23$ provide a useful basis for
studying how the competing intrinsic shapes affect the spin-isospin response of
this nucleus. Here we note that the prolate configuration for $^{56}$Ni is isolated by a high barrier compared to the oblate case.

Figure~\ref{fig:56Ni_gtmp} shows the calculated GT strength distributions in
$^{56}$Ni for both the $\mathrm{GT}^-$ and $\mathrm{GT}^+$ channels. In the
$\mathrm{GT}^-$ channel, where experimental data from the $(p,n)$ reaction are
available, the measured strength is distributed over a relatively broad
excitation-energy interval. By contrast, the spherical solution ($\beta_2=0$)
produces a strongly concentrated resonance around $E_{\mathrm{ex}}\approx5$ MeV,
indicating insufficient fragmentation compared with the experimental data. The deformed wave functions, on the other
hand, redistribute the strength over a wider energy range and generate a more
fragmented pattern. In particular, the oblate configuration enhances the
low-energy strength and yields a broader distribution, which appears to be
qualitatively closer to the experimental trend than the pure spherical
solution. This comparison suggests that the observed GT response in $^{56}$Ni
is not  consistent with a single rigid spherical configuration.

A similarly strong shape dependence is predicted in the $\mathrm{GT}^+$ channel,
which is directly relevant to electron capture on $^{56}$Ni. The spherical
solution again gives a dominant resonance near $E_{\mathrm{ex}}\approx5$ MeV,
whereas the prolate and oblate solutions produce stronger fragmentation and a
larger redistribution of strength toward lower and intermediate excitation
energies. Thus, an oblate shape or a shape coexistence between oblate and spherical configurations may manifest itself in the experimental GT strength distribution of  $^{56}$Ni though no  experimental
$\mathrm{GT}^+$ data are presently available to confirm or disprove this hypothesis.

The corresponding electron-capture rates are displayed in
Fig.~\ref{fig:56Ni_rate}. The rates increase smoothly as temperature and
density increase, and their overall behavior is much less structured than that of the
underlying GT$^+$ strength distributions. In fact, the calculated EC curves for
the competing shapes are nearly superposed on the scale of the figure. This
indicates that, once folded with the stellar phase-space factor, the capture
rates become primarily sensitive to the gross features of the low- and
intermediate-energy GT$^+$ strength rather than to its detailed fragmentation
pattern. In addition, the positive $Q_{EC}$ value for the electron-capture channel
makes the rates less threshold-limited, due to the EC continuum, than in nuclei with negative $Q_{EC}$. Consequently, the shape effect is much more pronounced in the
differential GT response than in the integrated electron-capture rate. In this
sense, $^{56}$Ni provides a clear example in which shape coexistence strongly
influences GT spectroscopy, while the associated stellar weak rates remain
comparatively robust.

The authors of Ref. \cite{Niu2012}
treated $^{56}$Ni as a spherical nucleus, and calculated the GT transition strengths by using various Skyrme parameter sets within the RPA plus particle-vibration coupling (PVC).
Good agreement with experiment is found, mainly driven by the PVC effect. Deformation could be seen as an 
adiabatic limit of PVC, using a zero excitation-energy limit of  the vibration mode. 
This may explain why 
both models give a two-peak structure of the GT strength. However, more investigation may be worth to compare the results in more depth.

%


\begin{figure} 
\includegraphics[width=0.65\linewidth]{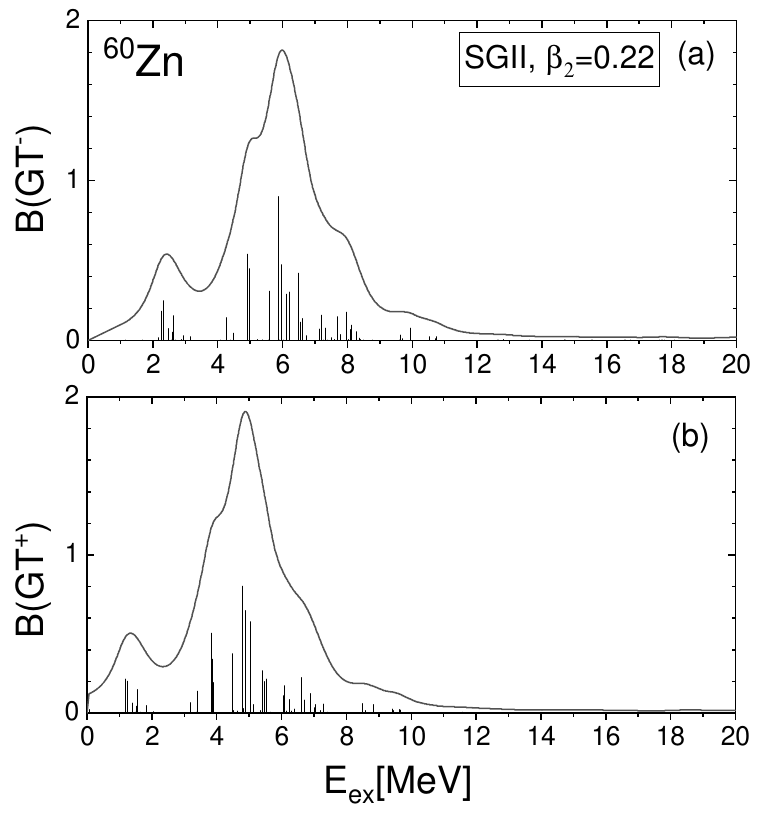}
\caption{{(Color online) Gamow--Teller $B(\mathrm{GT}^-)$ and $B(\mathrm{GT}^+)$ strength
distributions in $^{60}\mathrm{Zn}$ calculated with the SGII interaction for
the prolate shape $\beta_2=0.22$. Panels (a) and (b) show the
$\mathrm{GT}^-$ and $\mathrm{GT}^+$ channels, respectively.}}
\label{fig:60Zn_gtmp}
\end{figure}
\begin{figure} 
\includegraphics[width=0.75\linewidth]{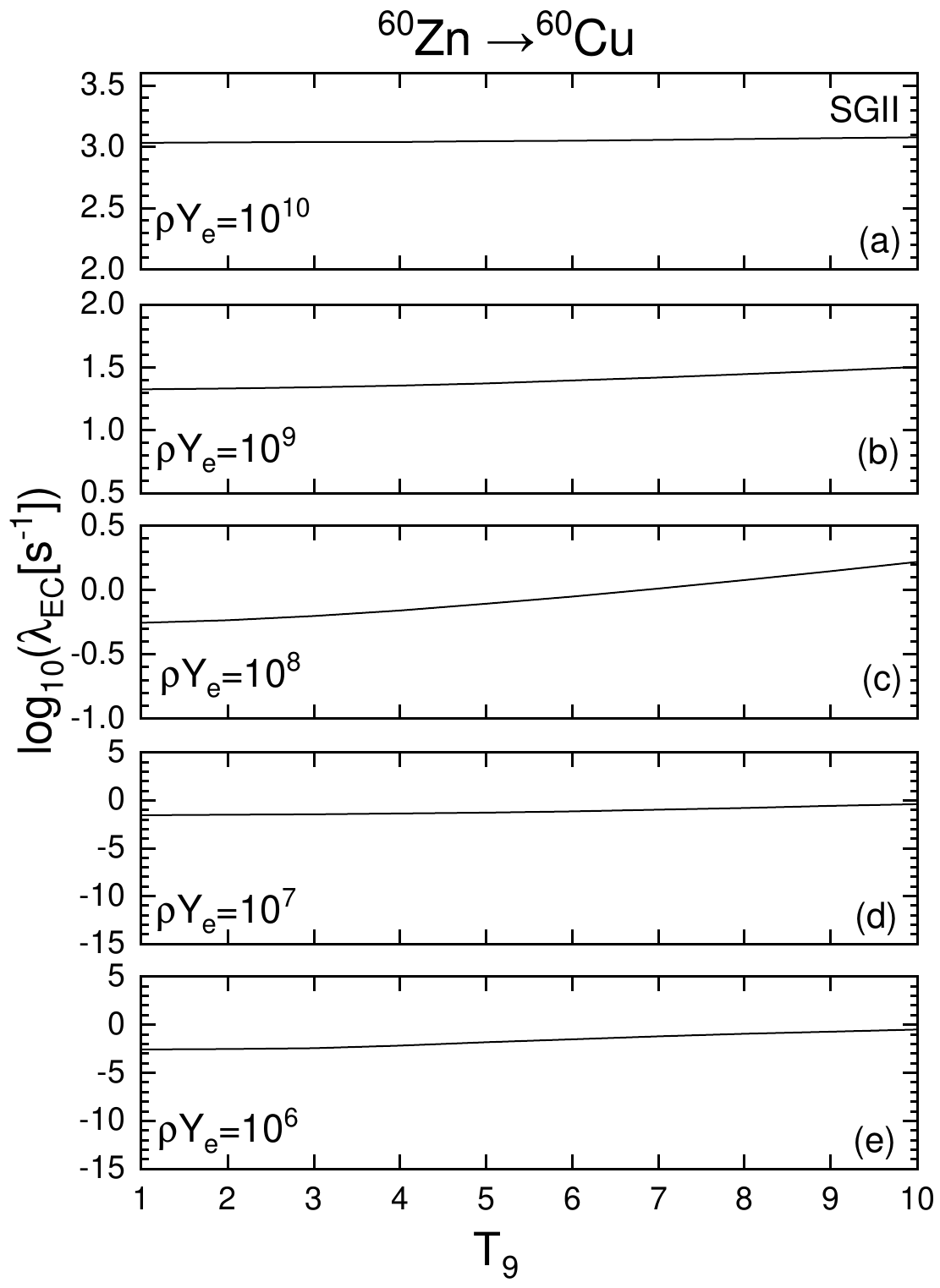}
\caption{{(Color online)  Electron-capture rates for $^{60}\mathrm{Zn}$ as a function of
temperature $T_9$ for densities $\rho Y_e=10^{10}, 10^9, 10^8, 10^7,$ and
$10^6$ mol/cm$^3$ in panels (a), (b), (c), (d), and (e), respectively. The
rates are calculated from the SGII GT$^+$ strength distribution shown in
Fig.~\ref{fig:60Zn_gtmp}.}}
\label{fig:60Zn_rate}
\end{figure}
%
\subsection{GT strengths and electron-capture rates in $^{60}$Zn}
\label{subsec:gt_ec_60zn}

The SGII potential-energy curve of $^{60}$Zn favors a well-developed prolate
minimum at $\beta_2\approx0.22$, while a shallower oblate minimum remains at a
higher energy. This suggests that, unlike the cases of  $^{56}$Ni and $^{64}$Ge, the low-energy structure of $^{60}$Zn is
dominated by the prolate configuration. Therefore, the present GT and
electron-capture calculations should be interpreted primarily as predictions for
the energetically favored prolate solution rather than as a full comparison
among nearly degenerate coexisting shapes.

Figure~\ref{fig:60Zn_gtmp} shows the calculated $\mathrm{GT}^-$ and
$\mathrm{GT}^+$ strength distributions for $^{60}$Zn obtained with the SGII
interaction at the prolate deformation $\beta_2=0.22$. Both channels exhibit a
broad and fragmented structure, including a low-energy shoulder around
$E_{\mathrm{ex}}\approx1$--$2$ MeV, a dominant collective component in the
intermediate region around $E_{\mathrm{ex}}\approx4$--$6$ MeV, and a long tail
extending to higher excitation energies. The GT$^+$ strength is concentrated
slightly lower in energy than the GT$^-$ strength, but the gross profiles of
the two channels have an  overall similarity, as expected for an $N=Z$ system with
nearly symmetric proton and neutron shell occupancies. These results indicate
that, in $^{60}$Zn, the static prolate deformation itself is sufficient to
generate a widely distributed GT response.

The corresponding electron-capture rates are displayed in
Fig.~\ref{fig:60Zn_rate}. The rates increase monotonically as functions of both temperature
and density. At high densities, the temperature dependence is rather weak,
showing that the large available electron phase space averages out the detailed
structure of the GT$^+$ distribution. At lower densities, the rates become more
temperature-dependent, reflecting the increasing importance of the low-lying
GT$^+$ strength in the case of the reduced phase-space availability. Nevertheless, the
overall behavior remains regular and robust, implying that for $^{60}$Zn the
stellar electron-capture rates are governed mainly by the gross features of
the low- and intermediate-energy GT$^+$ strength rather than by its detailed
peak-by-peak fragmentation.

Here, we note an interesting fact of the GT$^+$ strength among the three nuclei investigated. Although the gross GT$^+$ and GT$^-$ profiles are nearly symmetric in the
$N=Z$ nuclei, a small energy offset is visible in $^{60}$Zn, where the GT$^+$
strength is concentrated about 1 MeV lower than the GT$^-$ strength. This
difference should not be interpreted as a strong violation of the expected
GT$^\pm$ symmetry, but rather as a residual mirror-energy shift. Since the
horizontal axis in Fig.~7 is the daughter excitation energy $E_{\mathrm{ex}}$,
the bulk ground-state $Q$-value difference is largely absorbed into the
daughter mass offsets, and the remaining shift mainly reflects Coulomb
displacement and other small isospin-breaking effects in the mirror daughters
$^{60}$Cu and $^{60}$Ga. The reason why this offset is most visible in
$^{60}$Zn is probably due to the fact that its GT response is built on a single favored prolate
minimum and exhibits a relatively compact collective structure, so that the
residual mirror-energy difference is not washed out by strong fragmentation.
By contrast, in $^{56}$Ni and $^{64}$Ge the broader fragmentation and stronger
shape competition tend to mask such a small energy shift.
%


\begin{figure} 
\includegraphics[width=1.0\linewidth]{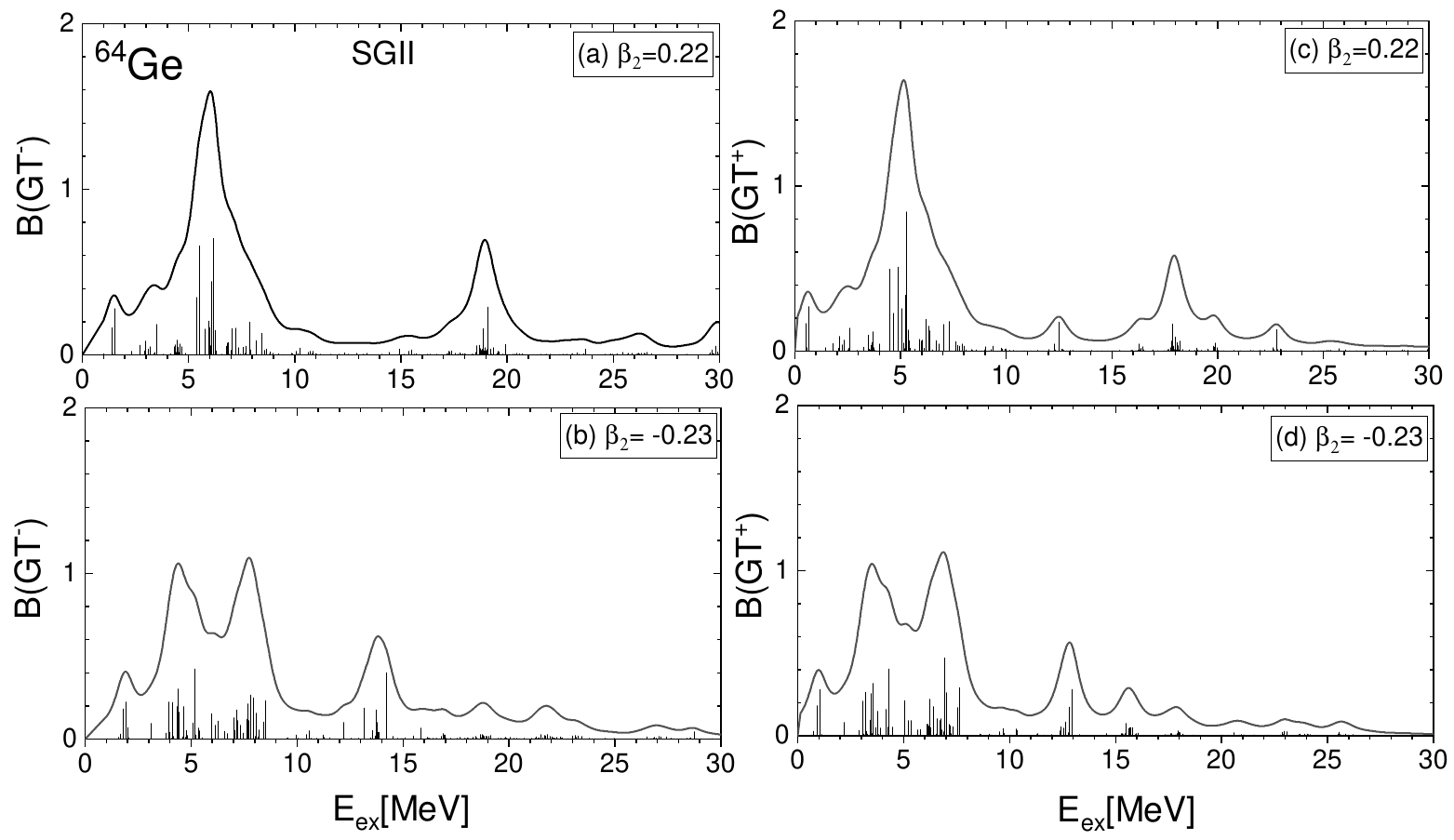}
\caption{(Color online) Gamow--Teller $B(\mathrm{GT}^-)$ and $B(\mathrm{GT}^+)$ strength
distributions in $^{64}\mathrm{Ge}$ calculated with the SGII interaction for
the prolate and oblate shapes. Panels (a) and (b) show the
$\mathrm{GT}^-$ strengths for $\beta_2=0.22$ and $\beta_2=-0.23$,
respectively, while panels (c) and (d) display the corresponding
$\mathrm{GT}^+$ strengths.
}
\label{fig:64Ge_gtmp}
\end{figure}
\begin{figure} 
\includegraphics[width=0.47\linewidth]{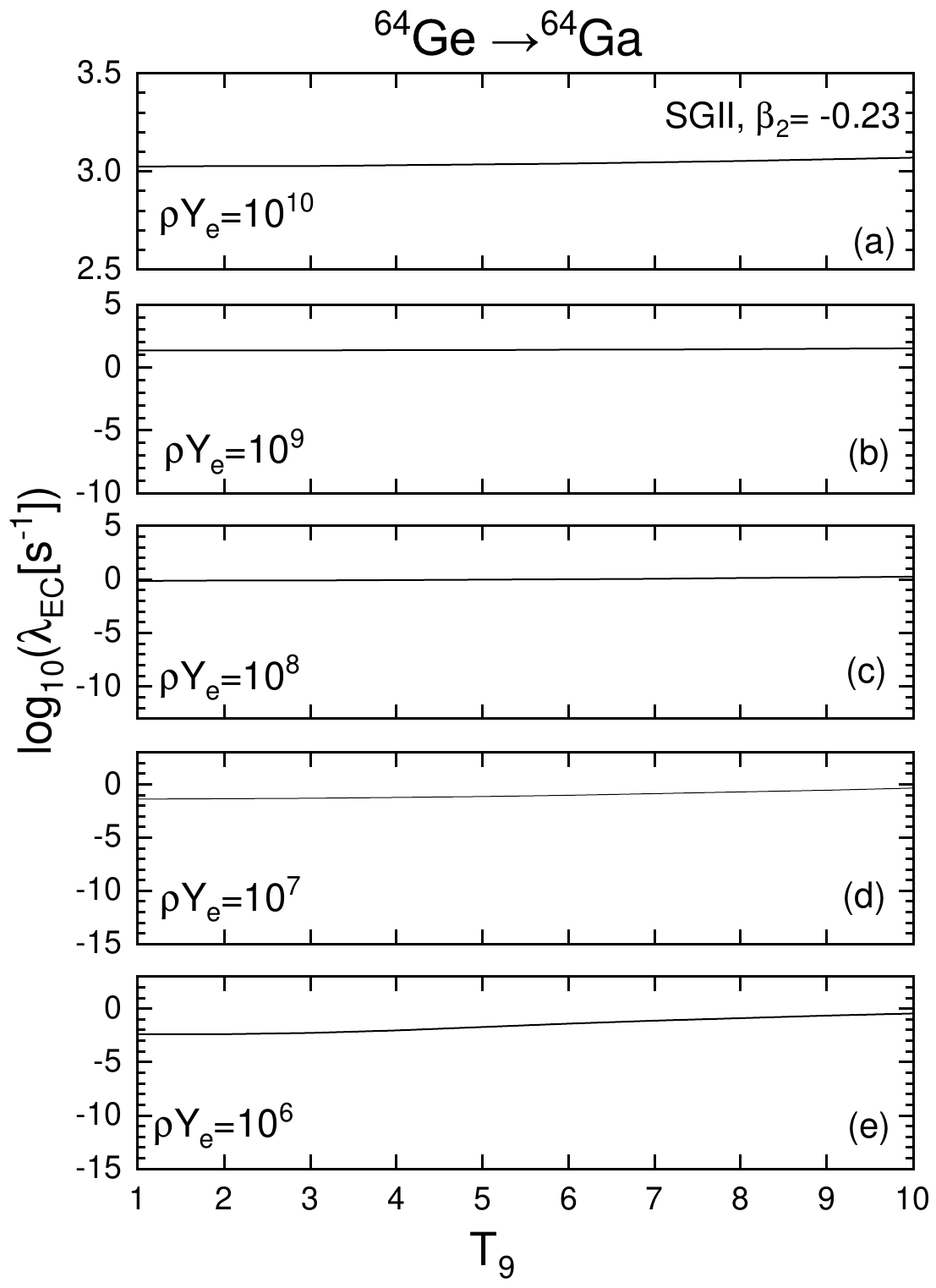}
\includegraphics[width=0.47\linewidth]{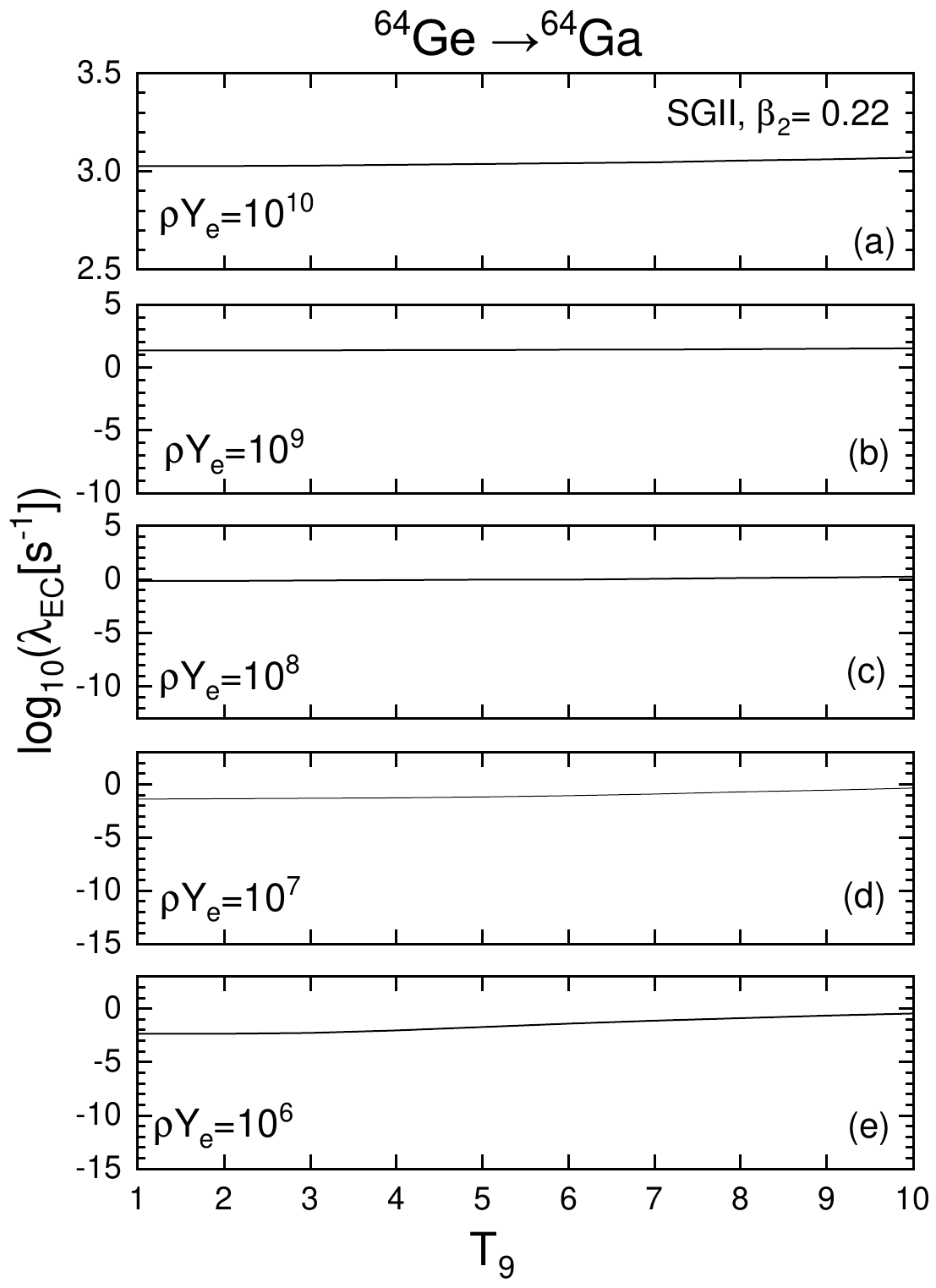}
\caption{(Color online) Electron-capture rates for $^{64}\mathrm{Ge}$ as a function of
temperature $T_9$ for densities $\rho Y_e=10^{10}, 10^9, 10^8, 10^7,$ and
$10^6$ mol/cm$^3$ in panels (a), (b), (c), (d), and (e), respectively. The
left and right columns correspond to the oblate and prolate solutions with
$\beta_2=-0.23$ and $\beta_2=0.22$, respectively.}
\label{fig:64Ge_rate}
\end{figure}
%

\subsection{GT strengths and electron-capture rates in $^{64}$Ge}
\label{subsec:gt_ec_64ge}

As seen in Fig. 1, the potential-energy curve of $^{64}$Ge,  exhibiting a pronounced competition
between prolate and oblate shapes, suggests that $^{64}$Ge should be interpreted as a possible 
oblate--prolate shape coexistence nucleus rather than as a system characterized by a single
rigid intrinsic shape, except for the height of the barrier, just like in $^{56}$Ni. Therefore, the comparison between the two constrained solutions provides a direct measure of how the competing deformations affect the
spin-isospin response. 
Here we also note that, given the height of the barrier and the other uncertainties, it is difficult to make a definite statement about shape coexistence. Generator Coordinate Method (GCM) 
calculations as well as the properties of the rotational band identified in the experiment, could be more efficient ways to identify shape coexistence \cite{Heyde2011}.  

Figure~\ref{fig:64Ge_gtmp} shows the calculated $\mathrm{GT}^-$ and
$\mathrm{GT}^+$ strength distributions for the prolate and oblate solutions. In
both channels, the prolate configuration ($\beta_2=0.22$) generates a dominant
collective low-energy resonance in the region $E_{\mathrm{ex}}\approx4$--$7$ MeV
together with an additional higher-lying component around
$E_{\mathrm{ex}}\approx18$--$19$ MeV. In contrast, the oblate configuration
($\beta_2=-0.23$) produces stronger fragmentation in the low-energy region and
shifts part of the higher-energy strength to lower excitation energies, leading
to a broader split distribution. Thus, the main deformation effect in
$^{64}$Ge is not a simple change in the total GT strength, but rather a
substantial redistribution of the resonance structure and fragmentation pattern.
The gross similarity between the $\mathrm{GT}^-$ and $\mathrm{GT}^+$ channels is
also consistent with the $N=Z$ character of this nucleus, although the detailed
peak positions and amplitudes remain shape dependent.

The corresponding electron-capture rates are displayed in
Fig.~\ref{fig:64Ge_rate}. Despite the clear shape dependence of the GT$^+$
strength distributions, the calculated rates for the oblate and prolate
solutions are nearly indistinguishable over the full range of temperatures and
densities considered here. This indicates that the stellar electron-capture
rates are governed mainly by the integrated low- and intermediate-energy
GT$^+$ strength that is common to both shapes, whereas the differences in the
detailed fragmentation pattern and in the higher-energy resonances are largely
washed out after phase-space folding. Therefore, $^{64}$Ge provides a clear
example in which shape coexistence is strongly reflected in GT spectroscopy but
has only a minor impact on the integrated stellar weak rates. From the
astrophysical point of view, this suggests that the effective weak lifetime of
$^{64}$Ge under the present stellar conditions is less sensitive to static
axial shape than the GT spectroscopy itself.

{\subsection{Electron chemical potential and critical density for electron capture}}

\begin{figure} 
\includegraphics[width=0.8\linewidth]{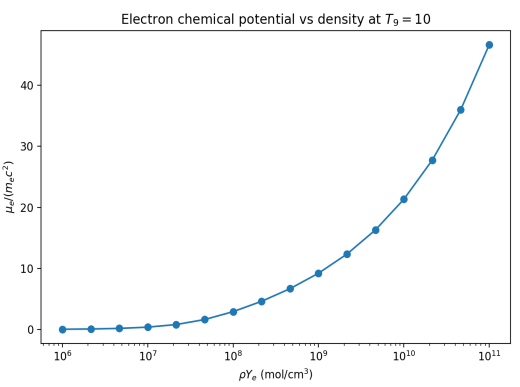}
\caption{(Color online){ Electron Chemical Potential $\mu_e$ in terms of $\rho Y_e$ at $T_9$ = 10}}
\label{fig:chemical_potential}
\end{figure}

The electron chemical potential $\mu_e$ in stellar matter is determined by the
density $\rho Y_e$ and temperature $T$ through the condition in Eq.~(\ref{eq:chemical_potential}). For a given temperature $T_9=10$, the numerical
solution of this equation shows that $\mu_e$ increases rapidly with density,
approximately following the degenerate scaling $\mu_e \propto (\rho Y_e)^{1/3}$, as shown in Fig.~\ref{fig:chemical_potential}. 
The choice $T_9 = 10$ is intended here only as a representative illustration of the density dependence of $\mu_e$.

This behavior plays a crucial role in determining the onset of electron capture
(EC) reactions. In particular, the EC rate is strongly enhanced when the
electron chemical potential becomes comparable to the relevant $Q$-value.

For $^{48}$Ti($e^-,\nu_e$)$^{48}$Sc, the electron capture $Q$-value from NNDC data is
$Q_{EC}(^{48}\mathrm{Ti}) = -3.990~\mathrm{MeV}$. Since $Q_{EC}<0$, the reaction requires energetic electrons from the ambient
plasma. The effective threshold condition is therefore $\mu_e \sim |Q_{EC}|$. Using the calculated $\mu_e(\rho Y_e)$ relation at $T_9=10$, we find that $\mu_e = 3.990~\mathrm{MeV}$ is reached at approximately
$\rho Y_e \approx 6.5\times 10^8~\mathrm{mol/cm^3}$. This density marks the onset of efficient electron capture on $^{48}$Ti.
Below this density, the EC rate is strongly suppressed due to the lack of
high-energy electrons, whereas above it the rate increases rapidly with
density, as shown in Fig.~{\ref{fig:48Ti_rate}.

For $^{56}$Ni($e^-,\nu_e$)$^{56}$Co, the NNDC value is $Q_{EC}(^{56}\mathrm{Ni}) = +2.132~\mathrm{MeV}$.
In this case, electron capture is energetically allowed even at low densities.
Thus, there is no strict threshold condition for EC. However, it is still
useful to define a characteristic density at which electron degeneracy begins
to significantly enhance the capture rate, namely, $\mu_e \sim Q_{EC}$. From the same $\mu_e(\rho Y_e)$ relation, we obtain $\mu_e = 2.132~\mathrm{MeV}$ at approximately $\rho Y_e \approx 1.8\times 10^8~\mathrm{mol/cm^3}.$

This density does not represent a strict threshold, but rather indicates the
onset of strong degeneracy effects that enhance the EC phase space.
The different behaviors of $^{48}$Ti and $^{56}$Ni arise from the sign of the
$Q$-value. For nuclei with negative $Q_{EC}$, such as $^{48}$Ti, electron
capture is a threshold process controlled by the electron chemical potential.
In contrast, for nuclei with positive $Q_{EC}$, such as $^{56}$Ni, electron
capture is intrinsically allowed and the density dependence enters primarily
through the enhancement of the electron phase space.

In both cases, the key controlling parameter is the electron chemical
potential, which governs the available phase space for the reaction through
the relation $E_\nu = Q_{if} + E_e$, and therefore determines the density dependence of stellar weak interaction
rates.

\section{Summary and Conclusions}
In this work, we investigated the deformation dependence of
Gamow--Teller (GT) strength distributions and stellar electron-capture (EC)
rates in the medium-mass nuclei $^{48}$Ti, $^{56}$Ni, $^{60}$Zn, and
$^{64}$Ge within a deformed Skyrme-based QRPA framework using the SGII
interaction. The calculated potential-energy curves indicate distinct
deformation scenarios in the nuclei considered here: $^{48}$Ti exhibits a
soft near-spherical/prolate landscape, $^{56}$Ni shows pronounced competition
among spherical, prolate, and oblate configurations, $^{60}$Zn is dominated by
a favored prolate minimum, and $^{64}$Ge displays strong oblate--prolate shape
coexistence.

The present results show that the GT strength distributions are strongly
affected by the intrinsic nuclear shape. In $^{48}$Ti, the moderate deformation
shifts a substantial part of the GT$^+$ strength to higher excitation energy,
while the EC rates remain relatively robust except at low temperature and low
density, where the rates become sensitive to the low-lying GT$^+$ strength.
In $^{56}$Ni, competing intrinsic shapes lead to markedly different GT
fragmentation patterns, whereas the corresponding EC rates are nearly unchanged
on the scale of the figure. In $^{60}$Zn, the energetically favored prolate
configuration already generates a broad and collective GT response, and the EC
rates are mainly governed by the gross low- and intermediate-energy
GT$^+$ strength. In $^{64}$Ge, the strong oblate--prolate coexistence produces
clearly different GT$^\pm$ distributions, but the corresponding EC rates remain
almost indistinguishable over the stellar temperature--density range considered
here.

A common conclusion emerging from all four nuclei is that deformation and shape
coexistence influence the GT response much more strongly than the
integrated stellar EC rates. Deformation changes the centroid energies,
resonance splitting, and fragmentation patterns of the GT distributions, but a
large part of this structural sensitivity is washed out once the rates are
folded with the stellar phase-space factor. As a result, the EC rates are
controlled mainly by the overall placement of the low- and intermediate-energy
GT$^+$ strength and become noticeably shape sensitive only in the regime of low
temperature and low density, where the contribution of low-lying transitions is
most critical. In this sense, GT spectroscopy provides a more sensitive probe
of intrinsic-shape competition than the EC rates themselves.

The present study also indicates that available charge-exchange data, such as
those for $^{48}$Ti and $^{56}$Ni, provide valuable benchmarks for assessing
the reliability of the calculated GT distributions. At the same time, the results
for the more proton-rich nuclei $^{60}$Zn and $^{64}$Ge demonstrate that
microscopic predictions remain essential in regions where experimental
information is scarce. This is particularly relevant for proton-rich nuclei
close to the $N=Z$ line, where EC rates may contribute to the weak-interaction
timescale under astrophysical conditions.

{Finally, we note several limitations of the present approach. The stellar EC
rates have been evaluated from GT$^+$ strengths built on the parent ground
state only, and possible contributions from thermally populated excited states
have not been included explicitly. In addition, the present calculations do
not include the isoscalar ($T=0$) pairing correlations and are based on a
hybrid framework in which the static mean field and the residual interaction
are not derived from a fully self-consistent EDF scheme. These effects may be
important in proton-rich nuclei near the $N=Z$ line and deserve further study.}

{Beyond these nuclear-structure developments, the present results also provide
useful microscopic input for future astrophysical applications. In particular,
updated electron-capture rates for $^{56}$Ni and neighboring proton-rich
nuclei should be incorporated into Type Ia supernova nucleosynthesis
calculations in order to quantify their impact on the production of iron-group
elements. Such calculations may help clarify the role of weak interactions in
distinguishing high-density near-Chandrasekhar-mass explosion models from
lower-density burning scenarios, such as white-dwarf mergers
\cite{Mori2016,Mori2018}.}

{A second important extension concerns proton-rich neutrino-driven ejecta in
core-collapse supernovae and hypernovae. The GT strength functions obtained
here for $^{56}$Ni, $^{60}$Zn, and $^{64}$Ge can serve as valuable
nuclear-structure input for future estimates of the corresponding $(n,p)$ and
charged-current neutrino-induced reaction cross sections. Combined with
astrophysical reaction-network calculations, such studies will help assess
whether the $\nu p$ process can overcome the bottlenecks at
$^{56}$Ni--$^{60}$Zn--$^{64}$Ge and contribute to the production of the
problematic p-nuclei $^{92,94}$Mo and $^{96,98}$Ru. In particular, it will be
important to examine this issue together with the possible influence of
Hoyle-state de-excitation in $^{12}$C, which may alter the $\alpha$-capture
flow and the availability of neutrons, protons, and $\alpha$ particles in the
relevant proton-rich environments
\cite{Fro2006,Sasaki2022,Jin2020,Bishop2022}.
Such astrophysical applications are beyond the scope of the present paper and
will be addressed in future work.}

\section*{Acknowledgement}
This work was supported by the National Research Foundation of Korea (Grant Nos. RS-2025-00513410 and RS-2024-00460031). The work of MKC is supported by the National Research Foundation of Korea (Grant Nos. RS-2021-NR060129 and RS-2025-16071941). 
The work of H.S. is supported by the JSPS Grant-in-Aid for Scientific Research (C) under Grant No. JP26K07079, and the Chinese Academy of Sciences (CAS) President's International Fellowship Initiative (PIFI, Grant No. 2024PVA0003\_Y1). 
{The work of TK is supported by the National Key R{\&}D Program of China (2022YFA1602401) and the National Natural Science Foundation of China (No. 12335009 {\&} 12435010).}

\appendix
\addcontentsline{toc}{section}{Appendices}
\section*{Appendices}
{\section{Phase-space factors for electron capture and $\beta^+$ decay}}

The weak transition rate can be written in the standard form
\begin{equation}
\lambda_{if}=\frac{\ln 2}{D} B_{if}\Phi_{if},
\end{equation}
where $B_{if}$ denotes the reduced transition strength for Fermi and
Gamow--Teller transitions, and $\Phi_{if}$ is the leptonic phase-space factor.

For electron capture,
\begin{equation}
e^-+(Z,A)_i \rightarrow (Z-1,A)_f+\nu_e,
\end{equation}
energy conservation gives
\begin{equation}
E_\nu = Q_{if}+E_e .
\end{equation}
The rate is obtained from Fermi's golden rule by integrating over the
initial electron distribution and the final neutrino phase space:
\begin{equation}
\lambda_{if}^{EC}\propto
\int p_e^2dp_e\, p_\nu^2dp_\nu\,
\delta(E_\nu-Q_{if}-E_e)\,
F(Z,E_e)\,S_e(E_e).
\end{equation}
After carrying out the neutrino integration with the delta function and
using $p_\nu c = E_\nu$ for a nearly massless neutrino, one obtains
\begin{equation}
\lambda_{if}^{EC}\propto
\int p_e^2dp_e\,
(Q_{if}+E_e)^2
F(Z,E_e)\,S_e(E_e).
\end{equation}
Introducing the dimensionless variables
\begin{equation}
\omega=\frac{E_e}{m_ec^2},\qquad
p=\sqrt{\omega^2-1},
\end{equation}
with $p_e^2dp_e \propto \omega p\, d\omega$, the phase-space factor becomes
\begin{equation}
\Phi_{if}^{EC}=
\int_{\omega_l}^{\infty}
\omega p (Q_{if}+\omega)^2
F(Z,\omega)\,S_e(\omega)\,d\omega .
\end{equation}
The lower limit is determined by the conditions $\omega\ge 1$ and
$Q_{if}+\omega\ge 0$, namely
\begin{equation}
\omega_l=
\begin{cases}
1, & Q_{if}>-1,\\
|Q_{if}|, & Q_{if}<-1.
\end{cases}
\end{equation}

For completeness, we also present the case of  $\beta^+$ decay, $(Z,A)_i \rightarrow (Z-1,A)_f+e^+ + \nu_e$, where energy conservation gives $E_\nu = Q_{if}-E_{e^+}$. The rate is then
\begin{equation}
\lambda_{if}^{\beta^+}\propto
\int p_e^2dp_e\, p_\nu^2dp_\nu\,
\delta(Q_{if}-E_e-E_\nu)\,
F(-Z,E_e).
\end{equation}
After eliminating the neutrino variable through the delta function,
\begin{equation}
\lambda_{if}^{\beta^+}\propto
\int p_e^2dp_e\,
(Q_{if}-E_e)^2
F(-Z,E_e).
\end{equation}
Using again the dimensionless electron energy $\omega$,
the corresponding phase-space factor becomes
\begin{equation}
\Phi_{if}^{\beta^+}=
\int_{1}^{Q_{if}}
\omega p (Q_{if}-\omega)^2
F(-Z,\omega)\,
\left[1-S_{e^+}(\omega)\right]
\left[1-S_\nu(Q_{if}-\omega)\right]
\,d\omega .
\end{equation}
In the absence of positron and neutrino blocking this reduces to
\begin{equation}
\Phi_{if}^{\beta^+}=
\int_{1}^{Q_{if}}
\omega p (Q_{if}-\omega)^2
F(-Z,\omega)\,d\omega .
\end{equation}

The difference between the two phase-space factors originates from the
different kinematics of the leptons. In electron capture, the incoming
electron contributes to the neutrino energy, giving rise to the factor
$(Q_{if}+\omega)^2$, whereas in $\beta^+$ decay the available decay
energy is shared between the outgoing positron and neutrino, leading to
the factor $(Q_{if}-\omega)^2$.

\end{document}